\documentclass[sn-nature,iicol]{sn-jnl}% Default with double column layout
\usepackage{graphicx}%
\usepackage{multirow}%
\usepackage{amsmath,amssymb,amsfonts}%
\usepackage{amsthm}%
\usepackage{mathrsfs}%
\usepackage[title]{appendix}%
\usepackage{textcomp}%
\usepackage{manyfoot}%
\usepackage{booktabs}%
\usepackage{algorithm}%
\usepackage{algorithmicx}%
\usepackage{algpseudocode}%
\usepackage{listings}%
\usepackage{pifont,xcolor} % for drawing triangles inside text
\usepackage{pdfpages}
\usepackage{pgffor}
\theoremstyle{thmstyleone}%
\usepackage[dvipsnames]{xcolor}
\usepackage{subfiles}
\usepackage{tikz}
\usepackage{pgfplots}
\usepgfplotslibrary{colorbrewer}
\pgfplotsset{compat=1.18}
\usetikzlibrary{positioning, calc, backgrounds, shapes.geometric, shapes.arrows, arrows.meta, shapes.symbols, decorations.text, fit}
\usetikzlibrary{external}
\usepgfplotslibrary{colormaps}
\usepgfplotslibrary{statistics}
\usepackage[most]{tcolorbox}
\usepackage{hyperref}
\usepackage{xcolor}
\usepackage{mdframed}
\newcommand{\entry}[1]{\textbf{\textsf{#1}}}
\newenvironment{marginnote}[1][]
{\begin{mdframed}[
  linewidth=0.8pt,
  roundcorner=4pt,
  linecolor=gray!60,
  backgroundcolor=gray!11,
  innertopmargin=3pt,
  innerbottommargin=3pt,
  innerleftmargin=8pt,
  innerrightmargin=8pt]
\small}
{\end{mdframed}}
\begin{document}
%%%%%%%%%%%%%%%%%%%%%%%%%%%%%%%%%%%%%%%%%%%%%%%%%%%%%%%%%%%%%%%%%%%%%%%%%%%%
\markboth{The Generative Science of Food Formulation}
{The Generative Science of Food Formulation}
%%%%%%%%%%%%%%%%%%%%%%%%%%%%%%%%%%%%%%%%%%%%%%%%%%%%%%%%%%%%%%%%%%%%%%%%%%%%
\title{\textbf{\textsf{\hspace*{3.cm}Artificial Intelligence \\ 
and the Generative Science of Food Formulation}}}
%%%%%%%%%%%%%%%%%%%%%%%%%%%%%%%%%%%%%%%%%%%%%%%%%%%%%%%%%%%%%%%%%%%%%%%%%%%%
\author[]{\fnm{Vahidullah} \sur{Tac}}\email{vtac@stanford.edu}
\author[]{\fnm{Ellen} \sur{Kuhl}}\email{ekuhl@stanford.edu}
%\author{Vahidullah Tac and Ellen Kuhl
\affil{Department of Mechanical Engineering, Stanford University, Stanford, California 94305, USA}
%%%%%%%%%%%%%%%%%%%%%%%%%%%%%%%%%%%%%%%%%%%%%%%%%%%%%%%%%%%%%%%%%%%%%%%%%%%%
\abstract{Food formulation requires balancing taste, nutrition, sustainability, and cost. Traditionally, new foods have emerged through empirical experimentation, expert intuition, and iterative refinement. \textcolor{black}{Now, artificial intelligence offers the opportunity to accelerate this process.} Yet despite rapid advances across food science, most AI applications remain isolated prediction and optimization tasks rather than parts of a broader scientific approach. Here we integrate these emerging technologies into a unified framework--the generative science of food formulation--in which digital food representations enable artificial intelligence to predict, discover, generate, organize, simulate, and optimize. We illustrate this approach through sustainability and nutrition, where generative artificial intelligence transforms environmental and nutritional metrics from post hoc evaluation criteria into explicit design objectives. Finally, we identify the data, models, benchmarks, and automation that will establish computational food design as a rigorous scientific discipline. Together, these advances have the potential to transform food formulation from an empirical discipline into a generative science.}
%%%%%%%%%%%%%%%%%%%%%%%%%%%%%%%%%%%%%%%%%%%%%%%%%%%%%%%%%%%%%%%%%%%%%%%%%%%%
%%%%%%%%%%%%%%%%%%%%%%%%%%%%%%%%%%%%%%%%%%%%%%%%%%%%%%%%%%%%%%%%%%%%%%%%%%%%
\maketitle
%%%%%%%%%%%%%%%%%%%%%%%%%%%%%%%%%%%%%%%%%%%%%%%%%%%%%%%%%%%%%%%%%%%%%%%%%%%%
%\tableofcontents
%%%%%%%%%%%%%%%%%%%%%%%%%%%%%%%%%%%%%%%%%%%%%%%%%%%%%%%%%%%%%%%%%%%%%%%%%%%%
%%%%%%%%%%%%%%%%%%%%%%%%%%%%%%%%%%%%%%%%%%%%%%%%%%%%%%%%%%%%%%%%%%%%%%%%%%%%
{\emph{Food formulation is becoming a generative science.}} 
Feeding a growing global population while improving human health and reducing environmental impact requires foods that simultaneously satisfy multiple competing objectives, including nutrition, sensory quality, sustainability, affordability, manufacturability, and consumer acceptance \citep{lappe71}. Every recipe therefore represents a complex compromise among biological, chemical, and physical constraints \citep{keesing22}. 
\textcolor{black}{Unlike most engineered systems, foods must balance these constraints with diverse human and societal requirements \citep{datta26}. This complexity creates an exceptionally high-dimensional design problem \citep{mcclements21a}.}
%%%%%%%%%%%%%%%%%%%%%%%%%%%%%%%%%%%%%%%%%%%%%%%%%%%%%%%%%%%%%%%%%%%%%%%%%%%%%
\begin{marginnote}[]
\entry{food formulation}
{Process of selecting ingredients, compositions, and processing conditions to achieve desired sensory, nutritional, functional, and sustainability objectives.}
\end{marginnote}
%%%%%%%%%%%%%%%%%%%%%%%%%%%%%%%%%%%%%%%%%%%%%%%%%%%%%%%%%%%%%%%%%%%%%%%%%%%%
%%%%%%%%%%%%%%%%%%%%%%%%%%%%%%%%%%%%%%%%%%%%%%%%%%%%%%%%%%%%%%%%%%%%%%%%%%%%%
\begin{marginnote}[]
\entry{generative science}
{Scientific paradigm that uses data, mechanistic models, and AI to predict, discover, and create new systems. In food science, it enables computa- tional food design.}
\end{marginnote}
%%%%%%%%%%%%%%%%%%%%%%%%%%%%%%%%%%%%%%%%%%%%%%%%%%%%%%%%%%%%%%%%%%%%%%%%%%%%
Traditionally, food scientists have navigated this massive design space through empirical experimentation, expert intuition, and iterative product development \citep{mcclements21}.
While this approach has produced remarkable innovations, it remains inherently slow and explores only a fraction of the enormous space of possible food formulations \citep{kuhl25}.

Artificial intelligence is beginning to transform this paradigm \citep{lecun15}. \textcolor{black}{In food science, artificial intelligence could span the entire design process}: predictive models \textcolor{black}{could} estimate food properties from ingredient composition, discovery algorithms \textcolor{black}{could} uncover scientific relationships hidden within experimental data \citep{schmidt09}, generative models \textcolor{black}{could} create entirely new recipes and formulations \citep{goodfellow14}, foundation models \textcolor{black}{could} organize knowledge across diverse food domains \citep{bommasani21}, emerging world models \textcolor{black}{could} simulate how foods evolve during processing and storage \citep{ha18}, and agentic AI systems \textcolor{black}{could}  support increasingly autonomous scientific workflows \citep{boiko23}. \textcolor{black}{Collectively, these advances could extend artificial intelligence from predicting the properties of existing foods toward designing entirely new formulations.}

More fundamentally, food science itself is becoming increasingly digital \citep{barabasi20}: ingredients, processing, nutrition, flavor, texture, sustainability, and shelf life can now be represented computationally through structured data that connect food composition to physical, chemical, sensory, and environmental properties \citep{datta26}. Once these digital representations become available, artificial intelligence can learn relationships between formulation and function, predict food performance before products exist, and ultimately generate new formulations that satisfy multiple competing objectives \citep{tac26}. 
%{\emph{Can food formulation become a generative science?}}
In much the same way that computer-aided design transformed engineering from analyzing existing structures to systematically creating new ones, artificial intelligence is transforming food formulation from evaluating existing recipes to computationally designing entirely new foods \citep{alsarayreh23}.

The convergence of digital food representations and modern artificial intelligence is beginning to reshape food science itself \citep{datta22}. For example, the food technology company NotCo has leveraged artificial intelligence to design an entirely new plant-based alternative to dairy milk using cabbage and pineapple \citep{pichara21}. As artificial intelligence evolves from prediction toward autonomous scientific discovery, food formulation is evolving from empirical recipe development toward the generative science of food formulation \citep{kuhl25}. Rather than relying primarily on trial-and-error experimentation, future food design will increasingly integrate digital representations, mechanistic understanding, predictive modeling, generative optimization, and autonomous experimentation within unified computational frameworks \citep{ghafarollahi25}.

In this review, we develop three complementary perspectives that define this emerging discipline: First, we describe the evolution of artificial intelligence from predictive models to automated discovery, generative models, foundation models, world models, and agentic AI. Second, we show how ingredients, nutrition, flavor, texture, sustainability, and shelf life become digital representations that artificial intelligence can predict, understand, generate, and optimize. Third, we illustrate how sustainability and nutrition become explicit design objectives that artificial intelligence can optimize alongside sensory quality and other constraints. Together, these perspectives establish a unified framework for computational food design and demonstrate how artificial intelligence is transforming food formulation from an empirical process into {\emph{the generative science of food formulation}}.
%%%%%%%%%%%%%%%%%%%%%%%%%%%%%%%%%%%%%%%%%%%%%%%%%%%%%%%%%%%%%%%%%%%%%%%%%%%%%
\begin{marginnote}[]
\entry{computational food design}
{Digital design of foods through predictive models, optimization, and artificial intelligence rather than trial-and-error experimentation.}
\end{marginnote}
%%%%%%%%%%%%%%%%%%%%%%%%%%%%%%%%%%%%%%%%%%%%%%%%%%%%%%%%%%%%%%%%%%%%%%%%%%%%
%%%%%%%%%%%%%%%%%%%%%%%%%%%%%%%%%%%%%%%%%%%%%%%%%%%%%%%%%%%%%%%%%%%%%%%%%%%%%%
\begin{figure*}[h]
\includegraphics[width=1.0\textwidth]{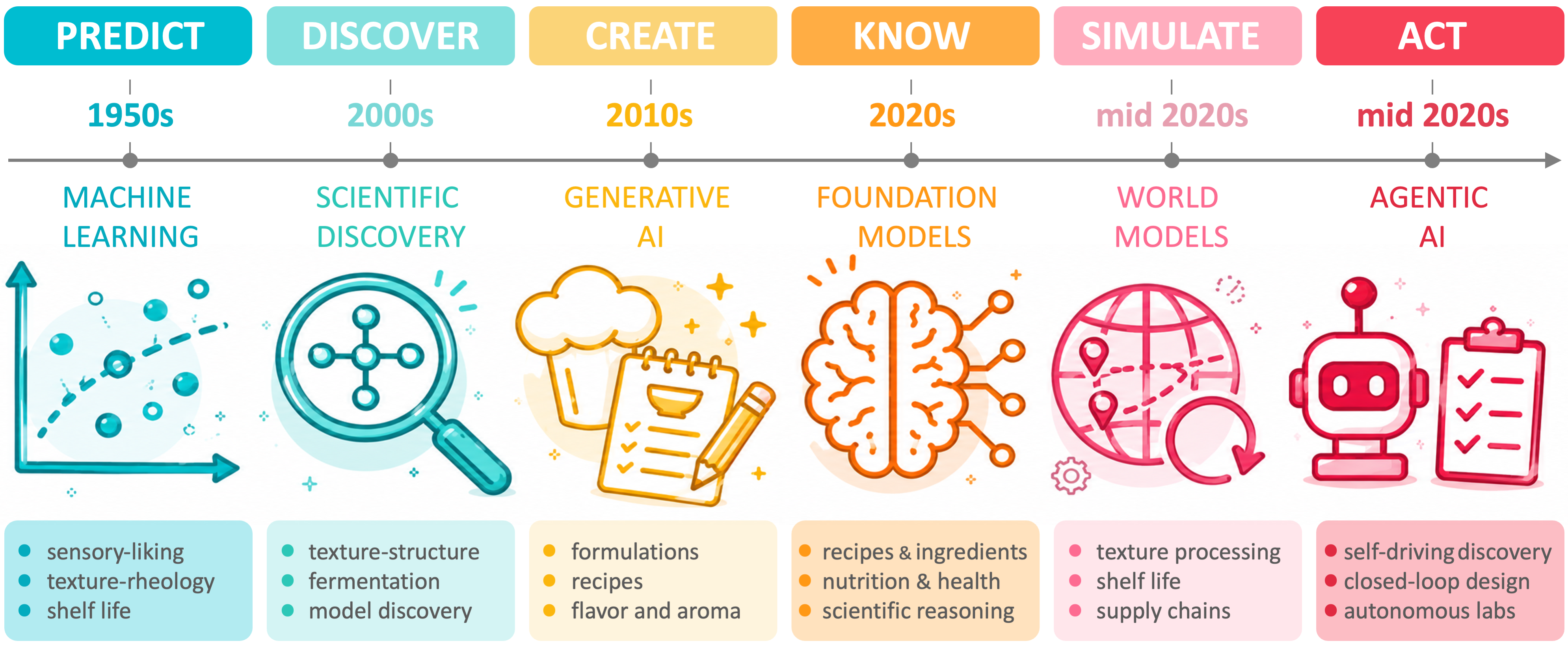}
\caption{{\bf{Evolution of artificial intelligence toward the generative science of food formulation.}} Artificial intelligence has evolved from prediction to scientific discovery, generation, knowledge integration, simulation, and increasingly autonomous action. \textcolor{black}{Rather than replacing one another, these six complementary capabilities increasingly converge within modern AI systems. Together, they transform AI} from predicting food properties to autonomously designing, evaluating, and refining new food formulations and establish the computational foundations of the generative science of food formulation.}
\label{fig01}
\end{figure*}
%%%%%%%%%%%%%%%%%%%%%%%%%%%%%%%%%%%%%%%%%%%%%%%%%%%%%%%%%%%%%%%%%%%%%%%%%%%%%%
%%%%%%%%%%%%%%%%%%%%%%%%%%%%%%%%%%%%%%%%%%%%%%%%%%%%%%%%%%%%%%%%%%%%%%%%%%%%
\section{THE EVOLUTION OF ARTIFICIAL INTELLIGENCE}
%%%%%%%%%%%%%%%%%%%%%%%%%%%%%%%%%%%%%%%%%%%%%%%%%%%%%%%%%%%%%%%%%%%%%%%%%%%%
Artificial intelligence has evolved from algorithms that {\emph{predict}} outcomes to systems that {\emph{discover}} scientific principles, {\emph{create}} new designs, {\emph{know}} and organize information, {\emph{simulate}} complex systems, and ultimately {\emph{act}} autonomously. These six complementary capabilities define the technological landscape of modern AI and provide the computational foundation for the next generation of AI-driven food formulation. Figure~\ref{fig01} summarizes this evolution.
%%%%%%%%%%%%%%%%%%%%%%%%%%%%%%%%%%%%%%%%%%%%%%%%%%%%%%%%%%%%%%%%%%%%%%%%%%%%
\subsection{Predictive AI}
%%%%%%%%%%%%%%%%%%%%%%%%%%%%%%%%%%%%%%%%%%%%%%%%%%%%%%%%%%%%%%%%%%%%%%%%%%%%
\begin{marginnote}[]
\entry{perceptron}
{Single-layer neural network that learns simple linear relationships between inputs and outputs from data.}
\end{marginnote}
%%%%%%%%%%%%%%%%%%%%%%%%%%%%%%%%%%%%%%%%%%%%%%%%%%%%%%%%%%%%%%%%%%%%%%%%%%%%
\begin{marginnote}[]
\entry{deep learning}
{Multilayer neural network that learns complex nonlinear relationships between inputs and outputs from data.}
\end{marginnote}
%%%%%%%%%%%%%%%%%%%%%%%%%%%%%%%%%%%%%%%%%%%%%%%%%%%%%%%%%%%%%%%%%%%%%%%%%%%%
Beginning in the 1950s, artificial intelligence first learned to \emph{predict} outcomes from data. Predictive AI represents the earliest and most established application of machine learning. The perceptron provides one of the first algorithms that was capable of learning input--output relationships directly from examples \citep{rosenblatt58}. Backpropagation then laid the foundation for training multilayer neural networks \citep{rumelhart86}. Deep learning now allows neural networks to capture increasingly complex nonlinear relationships across scientific domains \citep{lecun15}. In food science, predictive models \textcolor{black}{can} estimate sensory scores from ingredient composition, rheological properties from formulation, nutritional quality from recipes, and shelf life from packaging and storage conditions. Predictive AI transforms food development from empirical trial-and-error into data-driven prediction.
%%%%%%%%%%%%%%%%%%%%%%%%%%%%%%%%%%%%%%%%%%%%%%%%%%%%%%%%%%%%%%%%%%%%%%%%%%%%
\subsection{Discovery AI}
%%%%%%%%%%%%%%%%%%%%%%%%%%%%%%%%%%%%%%%%%%%%%%%%%%%%%%%%%%%%%%%%%%%%%%%%%%%%
Around the turn of the century, artificial intelligence began to \emph{discover} scientific principles from data. While predictive models identify relationships between inputs and outputs, discovery algorithms seek to uncover the underlying principles that govern these relationships. \textcolor{black}{Early approaches demonstrated this shift by recovering governing physical laws directly from experimental observations \citep{schmidt09}. Sparse identification later generalized this idea to nonlinear dynamical systems, where interpretable equations emerge from complex datasets \citep{brunton16}. 
\begin{marginnote}[]
\entry{automated model discovery}
{Autonomous identification of interpretable physics-based models and their parameters from data.}
\end{marginnote}
Constitutive neural networks now extend these concepts to automated model discovery, the autonomous identification of physics-based models and parameters directly from data \citep{linka23}. But rather than learning from data alone, these methods embed physical principles, symmetries, and constraints explicitly into the neural network architecture and inherently  ground their predictions in the physical world \citep{flaschel21}. In food science, physics-informed discovery-oriented AI can identify texture--structure relationships \citep{stpierre26b}, uncover drivers of consumer preference \citep{tac26a}, and reveal constitutive principles that govern food mechanics \citep{stpierre24}.} Discovery AI transforms data into scientific knowledge.
%%%%%%%%%%%%%%%%%%%%%%%%%%%%%%%%%%%%%%%%%%%%%%%%%%%%%%%%%%%%%%%%%%%%%%%%%%%%
\subsection{Generative AI}
%%%%%%%%%%%%%%%%%%%%%%%%%%%%%%%%%%%%%%%%%%%%%%%%%%%%%%%%%%%%%%%%%%%%%%%%%%%%
%%%%%%%%%%%%%%%%%%%%%%%%%%%%%%%%%%%%%%%%%%%%%%%%%%%%%%%%%%%%%%%%%%%%%%%%%%%%%
\begin{marginnote}[]
\entry{generative AI}
{Artificial intelligence that creates new data or designs from learned patterns. In food science, generative AI can generate entirely new recipes and ingredient formulations.}
\end{marginnote}
%%%%%%%%%%%%%%%%%%%%%%%%%%%%%%%%%%%%%%%%%%%%%%%%%%%%%%%%%%%%%%%%%%%%%%%%%%%%
The 2010s marked a major shift as artificial intelligence learned to \emph{create} new designs from data. Generative AI extends machine learning from analysis to synthesis. Variational autoencoders represent a probabilistic framework for learning latent representations and generating novel samples \citep{kingma14}. Generative adversarial networks demonstrate how models can create realistic new examples from existing data \citep{goodfellow14}. Diffusion models now achieve unprecedented performance across image, text, and scientific generation tasks \citep{ho20}. In food science, generative models \textcolor{black}{can} create new recipes, identify ingredient substitutions, propose flavor combinations, and design formulations that satisfy nutritional, sensory, or sustainability constraints \citep{tac26}. Generative AI explores possibilities that extend beyond existing products and datasets.
%%%%%%%%%%%%%%%%%%%%%%%%%%%%%%%%%%%%%%%%%%%%%%%%%%%%%%%%%%%%%%%%%%%%%%%%%%%%
\subsection{Foundation Models}
%%%%%%%%%%%%%%%%%%%%%%%%%%%%%%%%%%%%%%%%%%%%%%%%%%%%%%%%%%%%%%%%%%%%%%%%%%%%
%%%%%%%%%%%%%%%%%%%%%%%%%%%%%%%%%%%%%%%%%%%%%%%%%%%%%%%%%%%%%%%%%%%%%%%%%%%%%
\begin{marginnote}[]
\entry{foundation model}
{Large pretrained model that learns general representations from diverse data. In food science, it organizes recipes, nutrition, ingredients, and scientific knowledge.}
\end{marginnote}
%%%%%%%%%%%%%%%%%%%%%%%%%%%%%%%%%%%%%%%%%%%%%%%%%%%%%%%%%%%%%%%%%%%%%%%%%%%%
In the early 2020s, artificial intelligence began to \emph{know} more than any individual model before it. Foundation models represent a new paradigm in which large-scale pretraining produces general-purpose representations that capture knowledge across diverse domains. Transformer-based language models demonstrate how self-supervised learning can capture broad linguistic and semantic structure \citep{devlin19}. Large language models show that pretrained systems can perform diverse tasks with minimal task-specific training \citep{brown20}. The foundation-model framework formalizes these advances and highlights their broad scientific potential \citep{bommasani21}. In food science, foundation models \textcolor{black}{can} integrate recipes, nutrition databases, sensory studies, ingredient ontologies, and scientific literature into unified representations. Foundation models create connections across data modalities and disciplinary boundaries.
%%%%%%%%%%%%%%%%%%%%%%%%%%%%%%%%%%%%%%%%%%%%%%%%%%%%%%%%%%%%%%%%%%%%%%%%%%%%
\subsection{World Models}
%%%%%%%%%%%%%%%%%%%%%%%%%%%%%%%%%%%%%%%%%%%%%%%%%%%%%%%%%%%%%%%%%%%%%%%%%%%%
%%%%%%%%%%%%%%%%%%%%%%%%%%%%%%%%%%%%%%%%%%%%%%%%%%%%%%%%%%%%%%%%%%%%%%%%%%%%%
\begin{marginnote}[]
\entry{world model}
{AI model that predicts how complex systems evolve over time. In food science, it can simulate processing, fermentation, storage, and consumer response.}
\end{marginnote}
%%%%%%%%%%%%%%%%%%%%%%%%%%%%%%%%%%%%%%%%%%%%%%%%%%%%%%%%%%%%%%%%%%%%%%%%%%%%
More recently, artificial intelligence has begun to \emph{simulate} how complex systems evolve over time. Most predictive models estimate static outcomes. Instead, world models learn the dynamics that govern future states. Learned latent representations demonstrate how agents can construct internal models of dynamic environments \citep{ha18}. Latent dynamics models enable prediction and planning directly within learned representations \citep{hafner19}. Imagination-based learning shows how agents can learn effective behaviors through simulated interactions within virtual worlds \citep{hafner20}. In food science, world models offer a framework to simulate fermentation trajectories, texture evolution during processing, shelf-life degradation during storage, and interactions between formulation, manufacturing, and consumer behavior. World models enable virtual experimentation before physical testing begins.
%%%%%%%%%%%%%%%%%%%%%%%%%%%%%%%%%%%%%%%%%%%%%%%%%%%%%%%%%%%%%%%%%%%%%%%%%%%%
\subsection{Agentic AI}
%%%%%%%%%%%%%%%%%%%%%%%%%%%%%%%%%%%%%%%%%%%%%%%%%%%%%%%%%%%%%%%%%%%%%%%%%%%%
%%%%%%%%%%%%%%%%%%%%%%%%%%%%%%%%%%%%%%%%%%%%%%%%%%%%%%%%%%%%%%%%%%%%%%%%%%%%%
\begin{marginnote}[]
\entry{agentic AI}
{AI system that autonomously plans, uses tools, and executes tasks to achieve goals. In food science, it enables closed-loop experimentation and self-driving laboratories.}
\end{marginnote}
%%%%%%%%%%%%%%%%%%%%%%%%%%%%%%%%%%%%%%%%%%%%%%%%%%%%%%%%%%%%%%%%%%%%%%%%%%%%
The most recent development enables artificial intelligence to \emph{act} autonomously within scientific workflows. Agentic AI extends machine intelligence from reasoning to action. Reasoning-and-acting frameworks demonstrate how language models can combine deliberation with decision making \citep{yao23}. Tool-use architectures show that language models can learn to interact with external resources to solve complex tasks \citep{schick23}. Autonomous scientific systems demonstrate closed-loop research workflows in chemistry and other fields \citep{boiko23}. In food science, agentic systems may eventually formulate hypotheses, search the literature, design experiments, analyze results, and refine formulations within closed-loop discovery platforms. Agentic AI transforms artificial intelligence from a passive advisor into an active participant in scientific innovation. %\\[8.pt]
%%%%%%%%%%%%%%%%%%%%%%%%%%%%%%%%%%%%%%%%%%%%%%%%%%%%%%%%%%%%%%%%%%%%%%%%%%%%%%
\begin{figure*}[h]
\includegraphics[width=1.0\textwidth]{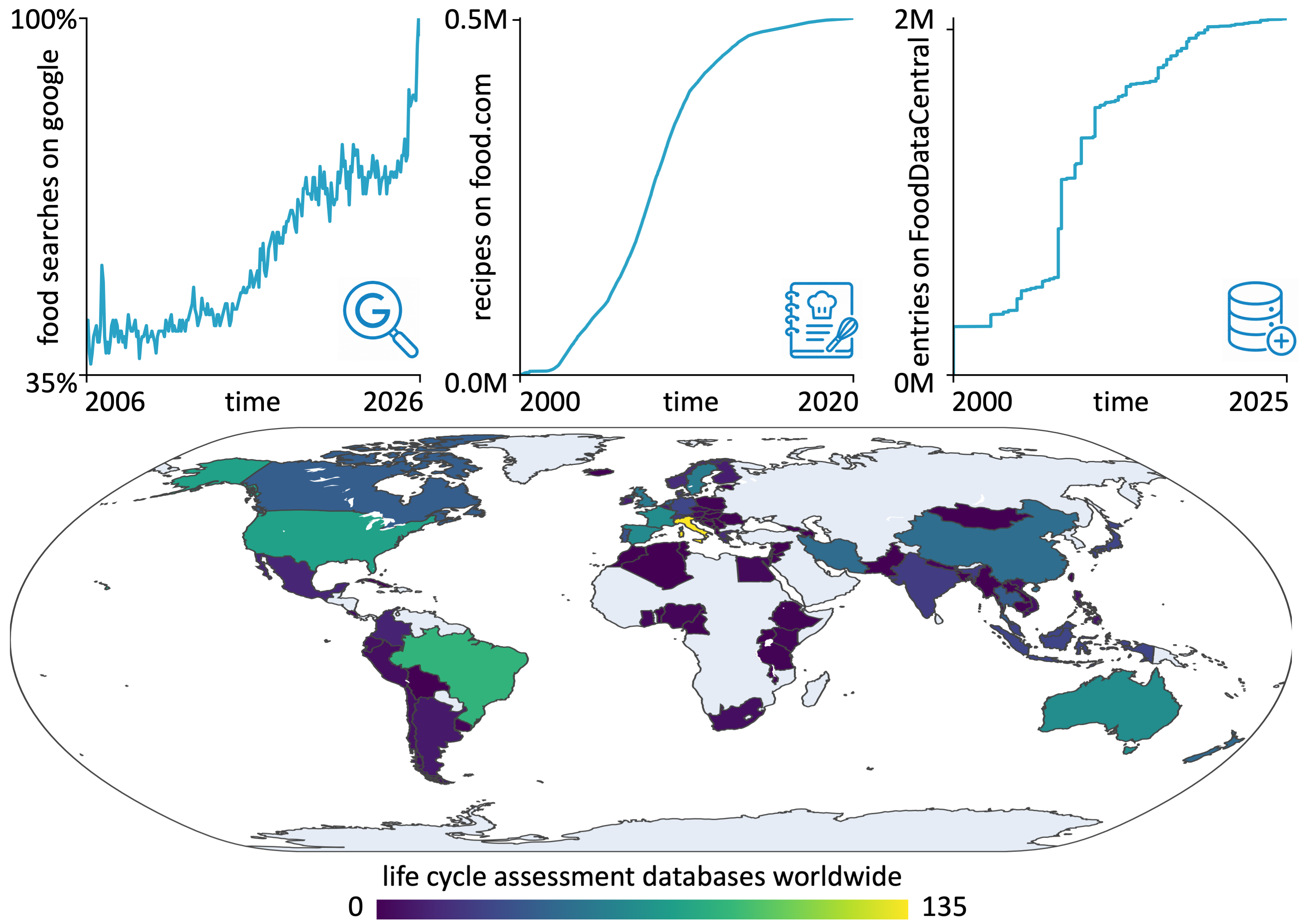}
\caption{{\bf{Expanding digital infrastructure for computational food design.}} Open food data are growing rapidly in scale and diversity. {\emph{Google}} searches for food are reaching an all-time high (top left), {\emph{Food.com}} contains more than 500,000 user-contributed recipes (top center), and {\emph{USDA FoodData Central}} now includes more than two million food entries (top right). Meanwhile, life cycle assessment databases cover agricultural products from more than 80 countries worldwide (bottom). Together, these digital resources provide the foundation for predictive, generative, and increasingly autonomous AI in food formulation.}
\label{fig02}
\end{figure*}\\[8.pt]
%%%%%%%%%%%%%%%%%%%%%%%%%%%%%%%%%%%%%%%%%%%%%%%%%%%%%%%%%%%%%%%%%%%%%%%%%%%%%%
\textcolor{black}{Taken together, these six complementary capabilities increasingly converge within modern AI systems rather than replace one another: predictive models estimate outcomes, discovery algorithms reveal mechanisms, generative models create candidates, foundation models organize knowledge, world models simulate future states, and agentic systems execute actions.} Together these six capabilities define the computational foundations of the generative science of food formulation. Figure~\ref{fig02} highlights the expanding digital food infrastructure that underpins this transformation. 
The following sections show how these complementary capabilities combine to transform food formulation: food representations provide the foundation for predictive and generative models, which increasingly optimize foods simultaneously for sensory quality, nutrition, and sustainability.
%%%%%%%%%%%%%%%%%%%%%%%%%%%%%%%%%%%%%%%%%%%%%%%%%%%%%%%%%%%%%%%%%%%%%%%%%%%%
\section{DATA FOUNDATIONS FOR AI-DRIVEN FOOD DESIGN}
\label{sec_data}
%%%%%%%%%%%%%%%%%%%%%%%%%%%%%%%%%%%%%%%%%%%%%%%%%%%%%%%%%%%%%%%%%%%%%%%%%%%%
Food becomes computationally designable only when we convert its formulation, \textcolor{black}{processing}, properties, and performance into structured digital representations. 
\textcolor{black}{Ingredients and composition, nutrition, flavor and taste, texture and rheology, and shelf life provide complementary digital representations of food formulation.} 
\textcolor{black}{Together, these representations define the {\emph{design variables}} that we can modify and the {\emph{design objectives}} that we can optimize, while food safety defines a non-negotiable {\emph{design constraint}} that every formulation must satisfy.} 
Just as computer-aided engineering relies on digital representations of geometry, materials, and performance, AI-driven food design relies on these representations to formalize its design space.
But unlike domains such as drug discovery and materials science, food science lacks centralized and standardized data infrastructures. Information remains fragmented across academic publications, proprietary industry databases, government nutrition surveys, sensory studies, and recipe repositories. This fragmentation creates both barriers and opportunities: barriers because predictive and generative models require diverse training data, and opportunities because integrating heterogeneous data sources can reveal relationships that remain invisible within individual datasets. 
%%%%%%%%%%%%%%%%%%%%%%%%%%%%%%%%%%%%%%%%%%%%%%%%%%%%%%%%%%%%%%%%%%%%%%%%%%%%%
\begin{marginnote}[]
\entry{design variable}
{\textcolor{black}{Quantity that an optimization algorithm {\emph{can}} modify. In food science, design variables include ingredients, compositions, and processing strategies.}}
\end{marginnote}
%%%%%%%%%%%%%%%%%%%%%%%%%%%%%%%%%%%%%%%%%%%%%%%%%%%%%%%%%%%%%%%%%%%%%%%%%%%%%
\textcolor{black}{Table~\ref{table01} summarizes these data foundations for computational food design.}
\textcolor{black}{The relevance and measurement of these representations naturally vary across food categories and processing contexts.}
The following sections examine five representative data modalities that illustrate the digital foundations of AI-driven food formulation.
%%%%%%%%%%%%%%%%%%%%%%%%%%%%%%%%%%%%%%%%%%%%%%%%%%%%%%%%%%%%%%%%%%%%%%%%%%%%%
\begin{marginnote}[]
\entry{design objective}
{\textcolor{black}{Criterion that an optimization algorithm {\emph{can}} maximize or minimize. In food science, objec\-tives include sensory quality, nutrition, sustain\-abi\-lity, manu\-factura\-bi\-lity, and cost.}}
\end{marginnote}
%%%%%%%%%%%%%%%%%%%%%%%%%%%%%%%%%%%%%%%%%%%%%%%%%%%%%%%%%%%%%%%%%%%%%%%%%%%%%
%%%%%%%%%%%%%%%%%%%%%%%%%%%%%%%%%%%%%%%%%%%%%%%%%%%%%%%%%%%%%%%%%%%%%%%%%%%%%
\begin{marginnote}[]
\entry{design constraint}
{\textcolor{black}{Requirement that every feasible solution {\emph{must}} satisfy. In food science, food safety defines a non-negotiable constraint.}}
\end{marginnote}
%%%%%%%%%%%%%%%%%%%%%%%%%%%%%%%%%%%%%%%%%%%%%%%%%%%%%%%%%%%%%%%%%%%%%%%%%%%%%
%%%%%%%%%%%%%%%%%%%%%%%%%%%%%%%%%%%%%%%%%%%%%%%%%%%%%%%%%%%%%%%%%%%%%%%%%%%%%
\begin{table*}[h]
\tabcolsep7.5pt
\caption{{\bfseries{Data foundations for AI-driven food design.}}
Digital food representations define the design variables artificial intelligence can modify, the design objectives it can optimize, and the design constraints it must satisfy.}
\label{table01}
%\begin{center}
%\begin{tabular}{@{}|p{2.4cm}|p{4.2cm}|p{4.6cm}|@{}}
\begin{tabular}{@{}|p{3.0cm}|p{6.0cm}|p{6.3cm}|@{}}
\hline
{\textsf{\bfseries{design variable}}} &
{\textsf{\bfseries{digital representation}}} &
{\textsf{\bfseries{representative data or metric}}}\\ \hline \hline
  ingredients \&  composition
& ingredient identity, quantity, composition, functionality
& FoodData Central, FooDB, FlavorDB, Food.com, recipe databases\\ \hline

  processing
& processing history and conditions
& temperature, time, pressure, shear, fermentation conditions\\
\hline \hline
{\textsf{\bfseries{design objective}}} &
{\textsf{\bfseries{digital representation}}} &
{\textsf{\bfseries{representative data or metric}}}\\ \hline \hline
  nutrition
& nutrient composition, quality, and bioavailability
& FoodData Central, HEI, Nutri-Score, PDCAAS, DIAAS\\ \hline

  flavor \&  taste
& chemical, sensory, and preference attributes
& FlavorDB, sensory panels, consumer studies\\ \hline

  texture \&  rheology
& mechanical and rheological properties
& mechanical testing, rheology, texture profile analysis \\ \hline

  shelf life \&  packaging
& stability and barrier properties
& storage stability, microbial growth, barrier measurements\\ \hline

  sustainability
& environmental impact metrics
& greenhouse gas emissions, land use, water use, eutrophication potential\\ \hline

  manufacturability
& processing and production requirements
& processability, equipment requirements, scale-up \\ \hline

  cost
& economic performance
& ingredient prices, processing costs, production costs \\
\hline \hline
{\textsf{\bfseries{design constraint}}} &
{\textsf{\bfseries{digital representation}}} &
{\textsf{\bfseries{representative data or metric}}}\\ \hline \hline
  food safety
& safety requirements and regulations
& microbial, chemical, and allergen limits \\ \hline

\end{tabular}
%\end{center}
\end{table*}
%%%%%%%%%%%%%%%%%%%%%%%%%%%%%%%%%%%%%%%%%%%%%%%%%%%%%%%%%%%%%%%%%%%%%%%%%%%%%
%%%%%%%%%%%%%%%%%%%%%%%%%%%%%%%%%%%%%%%%%%%%%%%%%%%%%%%%%%%%%%%%%%%%%%%%%%%%
\subsection{Ingredients and Composition} % design variables
%%%%%%%%%%%%%%%%%%%%%%%%%%%%%%%%%%%%%%%%%%%%%%%%%%%%%%%%%%%%%%%%%%%%%%%%%%%%
Ingredients define the fundamental design variables of food formulation. They determine nutritional composition, sensory properties, functionality, cost, and environmental impact. To become computationally designable, ingredients must first be represented digitally through chemical, nutritional, functional, and culinary descriptors.
%%%%%%%%%%%%%%%%%%%%%%%%%%%%%%%%%%%%%%%%%%%%%%%%%%%%%%%%%%%%%%%%%%%%%%%%%%%%%%
\begin{figure*}[h]
\includegraphics[width=1.0\textwidth]{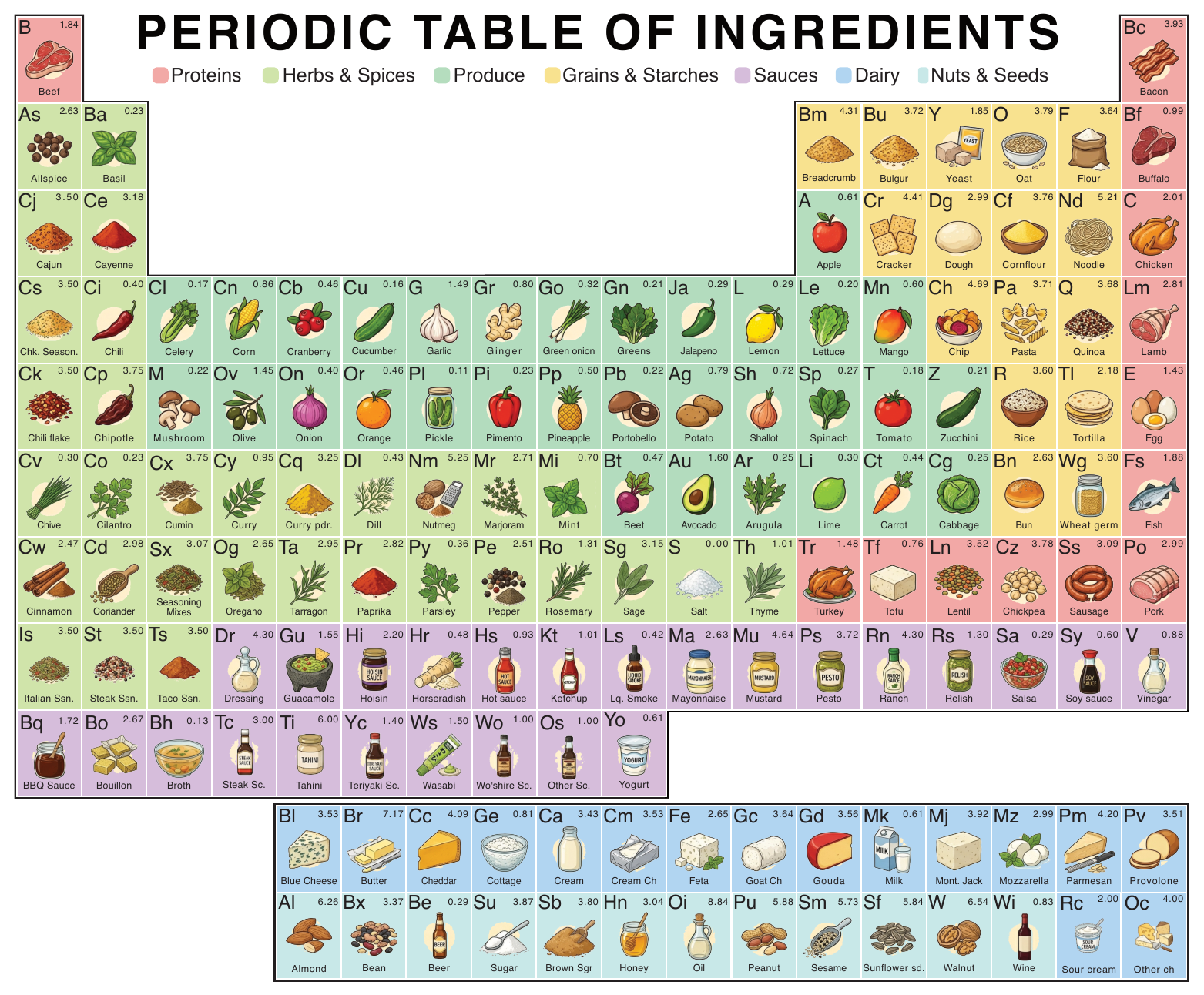}
\caption{{\bf{Periodic table of burger ingredients.}} The ingredient library consists of 146 ingredients from 2,216 burger recipes from Food.com and organizes them into ingredient families, including proteins, herbs and spices, produce, grains and starches, sauces, dairy, and nuts and seeds.
As the most fundamental digital representation of food, it defines the discrete formulation space from which generative AI learns ingredient relationships and creates new recipes.}
\label{fig03}
\end{figure*} %\\[8.pt]
%%%%%%%%%%%%%%%%%%%%%%%%%%%%%%%%%%%%%%%%%%%%%%%%%%%%%%%%%%%%%%%%%%%%%%%%%%%%%%
Ingredient representations originate from complementary chemical, nutritional, and culinary databases. {\emph{FooDB}} catalogs more than 70,000 chemical compounds across common foods and links molecular structure to taste, aroma, and bioactivity. {\emph{USDA FoodData Central}} provides nutrient profiles for more than 300,000 foods across 168 nutrient types and serves as a foundation for ingredient-level nutritional analysis \citep{ahuja20}. {\emph{FlavorDB}} maps over 25,000 flavor molecules across more than 900 foods and enables computational exploration of chemical flavor space \citep{ahn11}. Together, these resources represent ingredients at the levels of composition, functionality, and flavor.
\textcolor{black}{Ingredients differ not only in composition, but also in processing history and microstructure. Processing operations such as drying, extrusion, fermentation, thermal treatment, emulsification, and cell disruption modify food structure, functionality, digestibility, bioavailability, and sensory properties \citep{aguilera22}.}
Beyond individual ingredients, recipe datasets capture how ingredients combine into foods. The {\emph{Food.com}} dataset contains 522,517 recipes with 57,178 unique ingredient combinations and enables statistical learning of ingredient co-occurrence patterns \citep{food_com_recipes}. Recent work has shown that {\it{diffusion models}} 
%%%%%%%%%%%%%%%%%%%%%%%%%%%%%%%%%%%%%%%%%%%%%%%%%%%%%%%%%%%%%%%%%%%%%%%%%%%%%
\begin{marginnote}[]
\entry{diffusion model}
{Generative model that creates new samples by progressively removing noise. In food science, it generates novel recipes and ingredient formulations.}
\end{marginnote}
%%%%%%%%%%%%%%%%%%%%%%%%%%%%%%%%%%%%%%%%%%%%%%%%%%%%%%%%%%%%%%%%%%%%%%%%%%%%%
can learn the structure of ingredient combinations directly from recipe data and capture both ingredient popularity and higher-order correlations without explicit supervision \citep{tac26}. \textcolor{black}{Figure~\ref{fig03} illustrates the ingredient library for all burger formulation from {\emph{Food.com}}. As the most fundamental digital representation of food, it defines the discrete formulation space from which generative AI can learn and create entirely new burger recipes.}
%%%%%%%%%%%%%%%%%%%%%%%%%%%%%%%%%%%%%%%%%%%%%%%%%%%%%%%%%%%%%%%%%%%%%%%%%%%%%
\begin{marginnote}[]
\entry{formulation space}
{Set of possible ingredient combinations and quantities available for food design.}
\end{marginnote}
%%%%%%%%%%%%%%%%%%%%%%%%%%%%%%%%%%%%%%%%%%%%%%%%%%%%%%%%%%%%%%%%%%%%%%%%%%%%%
%%%%%%%%%%%%%%%%%%%%%%%%%%%%%%%%%%%%%%%%%%%%%%%%%%%%%%%%%%%%%%%%%%%%%%%%%%%%%
%\begin{marginnote}[]
%\entry{ontology}
%{Formal representation of concepts and their relationships within a domain. In food science, it organizes and integrates ingredients, nutrients, foods, and their properties.}
%\end{marginnote}
%%%%%%%%%%%%%%%%%%%%%%%%%%%%%%%%%%%%%%%%%%%%%%%%%%%%%%%%%%%%%%%%%%%%%%%%%%%%%
Despite recent progress, ingredient data remain fragmented across chemical, nutritional, sensory, and culinary domains. Government databases are largely open, whereas flavor and functionality data often remain proprietary. Molecular characterization requires specialized analytical workflows, and many studies report only aggregated results. Developing unified ingredient ontologies, standardized descriptors, and interoperable databases will be essential for connecting these disparate representations and enabling systematic AI-driven food formulation.
%%%%%%%%%%%%%%%%%%%%%%%%%%%%%%%%%%%%%%%%%%%%%%%%%%%%%%%%%%%%%%%%%%%%%%%%%%%%
\subsection{Nutrition} % health objectives
%%%%%%%%%%%%%%%%%%%%%%%%%%%%%%%%%%%%%%%%%%%%%%%%%%%%%%%%%%%%%%%%%%%%%%%%%%%%
Nutrition defines one of the primary objectives of food formulation. While ingredients define the design variables of a recipe, nutritional metrics quantify its potential contribution to human health. To become computationally actionable, nutritional quality must be represented in forms that algorithms can evaluate, compare, and optimize.
%%%%%%%%%%%%%%%%%%%%%%%%%%%%%%%%%%%%%%%%%%%%%%%%%%%%%%%%%%%%%%%%%%%%%%%%%%%%%
\begin{marginnote}[]
\entry{HEI}
{Healthy Eating Index. Dietary quality score that measures adherence to nutritional guidelines. In food science, it guides healthier food design.}
\end{marginnote}
%%%%%%%%%%%%%%%%%%%%%%%%%%%%%%%%%%%%%%%%%%%%%%%%%%%%%%%%%%%%%%%%%%%%%%%%%%%%
Nutritional representations build upon food composition databases and nutrient profiling systems. {\emph{USDA FoodData Central}} contains nutrient profiles for more than 300,000 foods and serves as the primary reference for nutritional calculations in the United States \citep{ahuja20}. The {\emph{Food Patterns Equivalents Database}} translates foods into dietary components and enables evaluation against dietary guidelines \citep{bowman20}. Together, these resources connect ingredient composition to macronutrients, micronutrients, fiber, and other health-relevant constituents.
Beyond individual nutrients, composite scoring systems provide higher-level representations of dietary quality. The {\emph{Healthy Eating Index}} (HEI)
evaluates adherence to federal dietary guidelines \citep{krebs18}; {\emph{Nutri-Score}} summarizes nutritional quality through a graded scoring system \citep{julia18}; and the {\emph{balanced hybrid nutrient density score}} combines nutrient profiling with food-group adequacy \citep{drewnowski19}. These metrics transform high-dimensional nutrient profiles into quantitative objectives that can be incorporated directly into predictive and generative models.
%
%%%%%%%%%%%%%%%%%%%%%%%%%%%%%%%%%%%%%%%%%%%%%%%%%%%%%%%%%%%%%%%%%%%%%%%%%%%%%
\begin{marginnote}[]
\entry{PDCAAS}
{Protein Digestibility-Corrected Amino Acid Score. Metric that combines amino acid composition with protein digestibility to evaluate protein quality.}
\end{marginnote}
%%%%%%%%%%%%%%%%%%%%%%%%%%%%%%%%%%%%%%%%%%%%%%%%%%%%%%%%%%%%%%%%%%%%%%%%%%%%%
%%%%%%%%%%%%%%%%%%%%%%%%%%%%%%%%%%%%%%%%%%%%%%%%%%%%%%%%%%%%%%%%%%%%%%%%%%%%%
\begin{marginnote}[]
\entry{DIAAS}
{Digestible Indispensable Amino Acid Score. Metric that evaluates protein quality from digestible essential amino acids.}
\end{marginnote}
%%%%%%%%%%%%%%%%%%%%%%%%%%%%%%%%%%%%%%%%%%%%%%%%%%%%%%%%%%%%%%%%%%%%%%%%%%%%%
%
\textcolor{black}{Protein quality provides another nutritional dimension \citep{loveday19}: the {\emph{Protein Digestibility-Corrected Amino Acid Score}} (PDCAAS) and {\emph{Digestible Indispensable Amino Acid Score}} (DIAAS) combine amino acid composition with protein digestibility to evaluate protein quality \citep{fao13}.
Together, these complementary metrics quantify different dimensions of nutritional quality, including dietary balance, micronutrient adequacy, and protein quality.}
Despite these advances, nutritional representations remain incomplete. Food composition databases rarely capture uncertainty, nutrient bioavailability, food matrix effects, or individual physiological variability. Developing automated ingredient-to-nutrient pipelines with uncertainty quantification 
%%%%%%%%%%%%%%%%%%%%%%%%%%%%%%%%%%%%%%%%%%%%%%%%%%%%%%%%%%%%%%%%%%%%%%%%%%%%%
\begin{marginnote}[]
\entry{uncertainty quantification}
{Estimation of confidence in model predictions. In food science, it identifies reliable predictions and highlights where additional data are needed.}
\end{marginnote}
%%%%%%%%%%%%%%%%%%%%%%%%%%%%%%%%%%%%%%%%%%%%%%%%%%%%%%%%%%%%%%%%%%%%%%%%%%%%
and standardized nutritional ontologies will enable more robust and personalized AI-driven food formulation \citep{linka25}.
%%%%%%%%%%%%%%%%%%%%%%%%%%%%%%%%%%%%%%%%%%%%%%%%%%%%%%%%%%%%%%%%%%%%%%%%%%%%
\subsection{Flavor and Taste} % preference signals
%%%%%%%%%%%%%%%%%%%%%%%%%%%%%%%%%%%%%%%%%%%%%%%%%%%%%%%%%%%%%%%%%%%%%%%%%%%%
Flavor captures the primary preference signals that determine consumer acceptance. It emerges from the integration of taste, aroma, appearance, and mouthfeel. Capturing this multisensory experience requires representations at multiple levels, from chemical composition to human perception and consumer preference \citep{spence15}. 
%%%%%%%%%%%%%%%%%%%%%%%%%%%%%%%%%%%%%%%%%%%%%%%%%%%%%%%%%%%%%%%%%%%%%%%%%%%%%
\begin{marginnote}[]
\entry{flavor}
{Multisensory perception that integrates taste, smell, vision, and touch. It determines the overall sensory experience of foods.}
\end{marginnote}
%%%%%%%%%%%%%%%%%%%%%%%%%%%%%%%%%%%%%%%%%%%%%%%%%%%%%%%%%%%%%%%%%%%%%%%%%%%%
Flavor representations span molecular databases, sensory lexicons, and consumer studies. At the molecular level, {\emph{FlavorDB}} catalogs more than 25,000 flavor molecules across over 900 foods and provides a foundation for linking chemical composition to sensory properties \citep{ahn11}. At the perceptual level, trained descriptive panels use standardized lexicons to quantify flavor attributes and create reproducible sensory benchmarks \citep{lawless98}. At the consumer level, large-scale preference studies provide representations of consumer acceptance and reveal how sensory attributes translate into purchasing and consumption decisions.
The {\emph{Taste of the Industry}} benchmarks represent the first large-scale open sensory datasets that systematically compare plant-based and animal meats \citep{vandenbedem26} and dairy and dairy-free products \citep{koosis26b} through blinded consumer testing. 
%%%%%%%%%%%%%%%%%%%%%%%%%%%%%%%%%%%%%%%%%%%%%%%%%%%%%%%%%%%%%%%%%%%%%%%%%%%%%
\begin{marginnote}[]
\entry{taste}
{Sensation perceived by taste receptors. It includes sweet, sour, salty, bitter, and umami.}
\end{marginnote}
%%%%%%%%%%%%%%%%%%%%%%%%%%%%%%%%%%%%%%%%%%%%%%%%%%%%%%%%%%%%%%%%%%%%%%%%%%%%
The study surveyed more than 2,500 consumers across 14 product categories and generated over 800,000 data points \citep{vandenbedem26}. Its results revealed that only a small number of categories—including chicken filets, nuggets, and burgers—achieved sensory parity between plant-based and animal products. In this dataset, {\emph{penalty analysis}} identified savoriness, moistness, and juiciness as dominant drivers of consumer acceptance. 
Controlled restaurant and dining-hall studies provide complementary sensory representations across diverse consumer populations. Sensory data may originate from trained panels and untrained consumer groups \citep{stpierre24}, direct paired comparisons of plant-based and animal products \citep{stpierre26}, homogeneous university dining-hall populations \citep{koosis26}, and heterogeneous urban restaurant cohorts \citep{tac26}. 
Despite these methodological differences, a consistent finding emerges: animal products continue to outperform plant-based alternatives across most food categories, although several categories now approach sensory parity.
%%%%%%%%%%%%%%%%%%%%%%%%%%%%%%%%%%%%%%%%%%%%%%%%%%%%%%%%%%%%%%%%%%%%%%%%%%%%%
\begin{marginnote}[]
\entry{benchmark}
{Standardized dataset for reproducible comparison of AI models.}
%{Standardized dataset used to compare algorithms fairly. In food science, benchmarks enable reproducible evaluation of predictive and generative AI models.}
\end{marginnote}
%%%%%%%%%%%%%%%%%%%%%%%%%%%%%%%%%%%%%%%%%%%%%%%%%%%%%%%%%%%%%%%%%%%%%%%%%%%%
Together, molecular databases, sensory panels, and consumer benchmarks form a hierarchy of flavor representations that connects chemistry to perception and preference, providing the foundation for predictive and generative models of food acceptance \citep{drake23}.
Consumer perception also reflects context, prior experience, expectations, associations, and multisensory cues that current datasets rarely capture comprehensively \citep{spence15}.
Moreover, sensory datasets remain fragmented across {\it{proprietary}} industry panels, small-scale academic studies, and inconsistent attribute definitions.
Establishing {\it{open}} sensory benchmarks with standardized protocols, large consumer cohorts, and paired instrumental–perceptual measurements will be essential for enabling rigorous cross-study comparisons and providing the training data required for AI models that predict consumer acceptance and generate foods people enjoy eating.
%%%%%%%%%%%%%%%%%%%%%%%%%%%%%%%%%%%%%%%%%%%%%%%%%%%%%%%%%%%%%%%%%%%%%%%%%%%%
\subsection{Texture and Rheology} % physical performance metrics
%%%%%%%%%%%%%%%%%%%%%%%%%%%%%%%%%%%%%%%%%%%%%%%%%%%%%%%%%%%%%%%%%%%%%%%%%%%%%%
Texture quantifies a physical representation of how food structure responds to deformation during consumption. Unlike flavor and nutrition, which rely on chemical and perceptual descriptors, texture is quantified through mechanical measurements that characterize how foods deform, fracture, recover, and flow. These measurements connect formulation and microstructure to mouthfeel and eating experience.
Texture representations arise from complementary mechanical testing methods. \textcolor{black}{One widely used method in food science is {\emph{texture profile analysis}} \citep{bourne78}}, 
%%%%%%%%%%%%%%%%%%%%%%%%%%%%%%%%%%%%%%%%%%%%%%%%%%%%%%%%%%%%%%%%%%%%%%%%%%%%%
\begin{marginnote}[]
\entry{texture profile analysis}
{Mechanical test that characterizes texture through repeated compression. In food science, it quantifies hardness, cohesiveness, springiness, and chewiness.}
\end{marginnote}
%%%%%%%%%%%%%%%%%%%%%%%%%%%%%%%%%%%%%%%%%%%%%%%%%%%%%%%%%%%%%%%%%%%%%%%%%%%%
a double-compression test that provides attributes such as hardness, cohesiveness, springiness, resilience, and chewiness \citep{dunne25}. Alternatively, {\it{rheological measurements}}
%%%%%%%%%%%%%%%%%%%%%%%%%%%%%%%%%%%%%%%%%%%%%%%%%%%%%%%%%%%%%%%%%%%%%%%%%%%%%
\begin{marginnote}[]
\entry{rheology}
{Science of material deformation and flow under applied forces. In food science, it characterizes food processing and texture.}
\end{marginnote}
%%%%%%%%%%%%%%%%%%%%%%%%%%%%%%%%%%%%%%%%%%%%%%%%%%%%%%%%%%%%%%%%%%%%%%%%%%%%
quantify elastic and viscous responses through storage and loss moduli \citep{boes26}. Together, these techniques generate quantitative descriptors that can be compared across ingredients, products, and formulations \citep{stpierre24a}.
Recent studies have established increasingly comprehensive open texture datasets for plant-based and animal foods \citep{stpierre24}. 
Yet instrumental measurements do not always translate directly into sensory perception \citep{vervenne25}. A recent study combining sensory evaluation and texture profile analysis across animal-, plant-, and mushroom-based burgers identified resilience as the strongest mechanical correlate of perceived meatiness and texture liking, while stiffness and hardness showed little relationship with consumer perception \citep{tac26a}. These findings illustrate that mechanical similarity does not necessarily imply sensory similarity, highlighting the challenge of linking instrumental texture measurements to human perception.
Developing mechanistic texture–perception models that integrate rheology, structure, oral processing, and sensory response remains a major challenge. Such multimodal representations 
%%%%%%%%%%%%%%%%%%%%%%%%%%%%%%%%%%%%%%%%%%%%%%%%%%%%%%%%%%%%%%%%%%%%%%%%%%%%%
\begin{marginnote}[]
\entry{multimodal representation}
{Unified representation that integrates multiple data types. In food science, it combines ingredients, nutrition, texture, flavor, sustainability, and consumer data.}
\end{marginnote}
%%%%%%%%%%%%%%%%%%%%%%%%%%%%%%%%%%%%%%%%%%%%%%%%%%%%%%%%%%%%%%%%%%%%%%%%%%%%
could bridge physical measurements and consumer perception, providing the foundation for predictive and generative models that optimize both texture and eating quality.
%%%%%%%%%%%%%%%%%%%%%%%%%%%%%%%%%%%%%%%%%%%%%%%%%%%%%%%%%%%%%%%%%%%%%%%%%%%%
\subsection{Shelf Life} % temporal performance metrics
%%%%%%%%%%%%%%%%%%%%%%%%%%%%%%%%%%%%%%%%%%%%%%%%%%%%%%%%%%%%%%%%%%%%%%%%%%%%
Shelf life measures the duration foods maintain acceptable quality during storage. It emerges from coupled interactions among microbial growth, chemical degradation, physical change, and packaging properties such as oxygen transmission, water vapor permeability, and light exposure \citep{robertson13}. Shelf-life representations therefore extend food design beyond formulation to include storage and distribution.
Shelf-life representations build upon predictive microbiology and packaging data. Predictive microbiology models such as {\emph{ComBase}} integrate more than 50,000 bacterial growth curves across 26,000 experimental conditions and support risk assessment and shelf-life estimation \citep{baranyi04}. Machine learning can predict spoilage from formulation, packaging, and storage conditions, but most existing models remain product-specific and rarely generalize across food categories. Recent reviews highlight the potential of AI to improve food safety through nontargeted screening \citep{wang25}, early warning systems, and emerging risk identification \citep{qian23}.
Compared with nutrition, flavor, and texture, shelf-life data remain relatively fragmented and underrepresented in AI-driven food design. Developing {\it{integrated datasets}} that link ingredient composition, processing history, packaging properties, storage conditions, and time-resolved quality measurements will enable predictive models that estimate shelf life across products and ultimately support the joint optimization of formulation and packaging for safety, quality, and sustainability. \\[8.pt]
\noindent Taken together, ingredients, nutrition, flavor, texture, and shelf life provide complementary representations of food systems. 
\textcolor{black}{Ingredients and processing define the design variables of formulation;} nutrition defines health objectives; flavor captures consumer preferences; texture quantifies physical performance; and shelf life measures temporal performance. Yet these representations remain fragmented across databases, disciplines, and experimental methodologies. Inconsistent measurement standards, limited interoperability, and restricted data access hinder reproducibility and model comparison.
Recent efforts have begun to establish {\emph{open benchmarks}} that combine formulation, sensory evaluation, mechanical characterization, nutritional analysis, and sustainability metrics \citep{dunne25,vandenbedem26,tac26}. The next generation of food datasets will need to integrate these modalities into unified representations that connect composition, structure, perception, function, and time. These unified representations provide the data foundation for AI models that predict food properties, discover structure--function relationships, and generate entirely new food formulations. Without them, AI remains confined to isolated tasks; with them, it can systematically navigate the complex trade-offs among taste, health, sustainability, and shelf life that define modern food systems. The next section turns from food representations to the predictive and generative models that learn from them.
%%%%%%%%%%%%%%%%%%%%%%%%%%%%%%%%%%%%%%%%%%%%%%%%%%%%%%%%%%%%%%%%%%%%%%%%%%%%
\section{AI FOR FOOD FORMULATION AND RECIPE DESIGN}
%%%%%%%%%%%%%%%%%%%%%%%%%%%%%%%%%%%%%%%%%%%%%%%%%%%%%%%%%%%%%%%%%%%%%%%%%%%%
Food formulation has traditionally relied on empirical trial-and-error, where recipes evolve through repeated experimentation and incremental refinement. Artificial intelligence replaces this paradigm with a computational design workflow that first learns the {\emph{forward mapping}} from formulation to function and then solves the {\emph{inverse problem}} of discovering formulations that satisfy desired objectives. This progression mirrors established paradigms in computational mechanics, materials science, and drug discovery. Predictive models estimate the properties of foods; generative models create new formulations; together they establish the computational foundation for AI-driven food design.
%%%%%%%%%%%%%%%%%%%%%%%%%%%%%%%%%%%%%%%%%%%%%%%%%%%%%%%%%%%%%%%%%%%%%%%%%%%%
\subsection{Predictive Modeling}
%%%%%%%%%%%%%%%%%%%%%%%%%%%%%%%%%%%%%%%%%%%%%%%%%%%%%%%%%%%%%%%%%%%%%%%%%%%%
Predictive models solve the {\emph{forward problem}} of food formulation.
Given ingredients, composition, processing conditions, or packaging, they estimate nutritional quality, texture, sensory perception, shelf life, or other measurable features without explicitly generating new recipes \citep{kuhl25}. Predictive models have become indispensable throughout the food value chain, from accelerating product development to reducing experimental costs and identifying promising formulations before physical prototypes are produced \citep{zhang25}. Learning accurate forward models represents the first step toward computational food design because prediction provides the foundation for optimization.
%%%%%%%%%%%%%%%%%%%%%%%%%%%%%%%%%%%%%%%%%%%%%%%%%%%%%%%%%%%%%%%%%%%%%%%%%%%%%
\begin{marginnote}[]
\entry{forward problem}
{Predicting system behavior from known inputs. In food science, it predicts nutrition, texture, flavor, or sustainability from food composition.}
\end{marginnote}
%%%%%%%%%%%%%%%%%%%%%%%%%%%%%%%%%%%%%%%%%%%%%%%%%%%%%%%%%%%%%%%%%%%%%%%%%%%%

One of the earliest applications of artificial intelligence in food science focused on predicting nutritional properties from ingredient composition. \textcolor{black}{Regression models and, more recently, machine learning algorithms estimate calories, macronutrients, micronutrients, glycemic response, and dietary quality indices directly from recipes and food composition databases using the nutritional representations introduced in Section \ref{sec_data} \citep{tac26}. But ingredient composition provides only part of the nutritional picture \citep{mcclements15}. Food processing, microstructure, bioavailability, and the food matrix also influence nutritional quality and physiological response \citep{capuano21}. These models support personalized nutrition, dietary recommendation systems, and rapid screening of alternative formulations without laboratory measurements. As food composition databases expand, future nutritional models should increasingly integrate processing history and food structure alongside ingredient composition \citep{norton14}.}

Texture remains one of the defining characteristics of food quality and a major determinant of consumer acceptance. Texture is fundamentally a {\emph{sensory property}} that reflects the structural, mechanical, and surface characteristics of foods and ultimately emerges through human perception \citep{szczesniak02}. Instrumental methods such as {\emph{texture profile analysis}}, {\emph{rheology}}, and {\emph{mechanical testing}} quantify these underlying physical properties and provide reproducible descriptors of stiffness, hardness, chewiness, cohesiveness, and viscoelasticity. \textcolor{black}{These physical representations provide high-quality training data for predictive AI models \citep{stpierre23}, particularly for alternative proteins, where reproducing the eating experience of conventional meat remains a major scientific challenge \citep{grossmann21,mcclements24}.}

\textcolor{black}{A particularly important recent development integrates {\emph{physics-based characterization}} with {\emph{sensory evaluation}}. Recent studies demonstrate statistically significant relationships between instrumental measurements and perceived texture \citep{tac26a}. Mechanical stiffness, rheological properties, and texture profile analysis correlate with sensory attributes such as softness, brittleness, moistness, meatiness, and overall acceptance, establishing quantitative links between measurable material properties and human perception \citep{stpierre26a}. Recent work also suggests that instrumental texture represents only one component of sensory perception. Large differences in mechanical properties do not necessarily translate into differences in consumer acceptance, particularly when product structure and oral processing alter texture perception \citep{koosis26a}. Together, these findings define both the relationships and the limitations that predictive AI faces when linking physical representations to sensory outcomes.}

Predictive modeling extends beyond formulation to estimate how foods evolve during storage and distribution. Models integrate ingredient composition, processing history, packaging, and environmental conditions to predict microbial growth, food safety risks, oxidation, moisture migration, texture degradation, color changes, nutrient losses, and overall quality retention \citep{deng21}. Traditional {\emph{kinetic models}} remain highly successful for describing many degradation processes \citep{vanboekel08,corradini18}, while machine learning increasingly complements these approaches by integrating heterogeneous experimental and environmental data. Together, these methods support packaging design, optimize storage conditions, improve inventory management, and reduce food waste across the food supply chain \citep{krupitzer24}. \\[8.pt]
Taken together, predictive models learn the mapping from {\emph{food representations to food properties}}: ingredients map onto nutritional profiles, mechanical properties, sensory responses, and shelf-life predictions. These {\emph{forward models}} transform the digital representations introduced in Section \ref{sec_data} into quantitative predictions that enable computational food design. Once AI can accurately predict the properties of a formulation, the problem naturally reverses: given a desired combination of taste, nutrition, sustainability, texture, cost, and manufacturability, which formulation should we create? This {\emph{inverse problem}} defines the central challenge of {\emph{generative food design}}.
%%%%%%%%%%%%%%%%%%%%%%%%%%%%%%%%%%%%%%%%%%%%%%%%%%%%%%%%%%%%%%%%%%%%%%%%%%%%
\subsection{Generative Modeling}
%%%%%%%%%%%%%%%%%%%%%%%%%%%%%%%%%%%%%%%%%%%%%%%%%%%%%%%%%%%%%%%%%%%%%%%%%%%%
Generative artificial intelligence extends food AI from {\emph{prediction}} to {\emph{creation}}. 
While predictive models solve the {\emph{forward problem}} by estimating the properties of existing formulations, generative models solve the {\emph{inverse problem}} by synthesizing entirely new recipes, ingredient combinations, and food concepts that have never existed before. 
\textcolor{black}{Rather than asking how a given recipe will perform, these models ask which recipe will best satisfy {\emph{multiple design objectives}}, including sensory quality, cost, nutrition, sustainability, and manufacturability \citep{zhang19}. Unlike these objectives, food safety defines a non-negotiable {\emph{design constraint}}. Future generative models should therefore satisfy microbiological, toxicological, allergen, and regulatory safety requirements {\it{before}} optimizing any other design objective \citep{qian23}.
Together, these advances transform food formulation from evaluating candidate designs into systematically exploring ingredient compositions and processing strategies that satisfy multiple competing design objectives and constraints.}
%%%%%%%%%%%%%%%%%%%%%%%%%%%%%%%%%%%%%%%%%%%%%%%%%%%%%%%%%%%%%%%%%%%%%%%%%%%%%
\begin{marginnote}[]
\entry{inverse problem}
{Identifying inputs that produce desired system behavior. In food science, it designs recipes that satisfy nutritional, sensory, and sustainability objectives.}
\end{marginnote}
%%%%%%%%%%%%%%%%%%%%%%%%%%%%%%%%%%%%%%%%%%%%%%%%%%%%%%%%%%%%%%%%%%%%%%%%%%%%

Early computational studies revealed that palatable recipes occupy highly structured regions within ingredient space rather than arbitrary collections of ingredients. The {\emph{Flavor Network}} demonstrated that ingredient combinations follow statistical regularities in shared flavor compounds and culinary traditions, which means recipes form navigable networks rather than isolated formulations \citep{ahn11}. Later, {\emph{Recipe1M}} learned joint representations of ingredients, images, and cooking instructions, allowing AI systems to organize recipes within continuous latent spaces that capture similarities in both composition and preparation \citep{recipe1m}. 
These learned representations transformed recipes into structured design objects that generative models can navigate, interpolate, and ultimately create.
%%%%%%%%%%%%%%%%%%%%%%%%%%%%%%%%%%%%%%%%%%%%%%%%%%%%%%%%%%%%%%%%%%%%%%%%%%%%%
\begin{marginnote}[]
\entry{latent space}
{Continuous mathematical representation learned from data that captures underlying similarities. In food science, it organizes recipes and formulations for generation and optimization.}
\end{marginnote}
%%%%%%%%%%%%%%%%%%%%%%%%%%%%%%%%%%%%%%%%%%%%%%%%%%%%%%%%%%%%%%%%%%%%%%%%%%%%%

{\emph{Large language models}} naturally extend this idea because recipes represent food through text: structured ingredient lists, natural-language cooking instructions, and user-defined design goals all exist in textual form \citep{brown20}. Users can prompt these models with dietary preferences, nutritional goals, available ingredients, or cultural styles, and the models produce coherent recipes together with cooking instructions, ingredient substitutions, and interactive explanations of ingredient choices \citep{zhou25}. They summarize scientific knowledge, suggest modifications, and refine recipes through conversational interaction, making foundation models powerful assistants for recipe development, scientific reasoning, knowledge integration, and personalization \citep{thomas25}.
%%%%%%%%%%%%%%%%%%%%%%%%%%%%%%%%%%%%%%%%%%%%%%%%%%%%%%%%%%%%%%%%%%%%%%%%%%%%%
\begin{marginnote}[]
\entry{large language model}
{Foundation model trained to understand and generate natural language. In food science, it organizes recipes, nutritional knowledge, and scientific literature.}
\end{marginnote}
%%%%%%%%%%%%%%%%%%%%%%%%%%%%%%%%%%%%%%%%%%%%%%%%%%%%%%%%%%%%%%%%%%%%%%%%%%%%%
However, large language models primarily generate statistically plausible text. They predict the next token from previously observed text and therefore reproduce patterns that exist in their training data. Prompt engineering and reasoning frameworks improve consistency, but these models do not directly optimize recipes for measurable objectives such as consumer acceptance, production cost, nutritional quality, or greenhouse gas emissions. Their greatest strength lies in representing and communicating culinary knowledge rather than optimizing formulation space.

{\emph{Diffusion models}} shift the focus from language generation to {\emph{food design}}. Instead of predicting text, these models learn probability distributions over recipes and sample entirely new formulations from these learned distributions. Food formulations combine discrete ingredient selection with continuous ingredient quantities, so recipes form {\emph{hybrid discrete--continuous design objects}} that require specialized generative architectures. 
%\textcolor{black}{Throughout this review, we use burgers as a model system because their modular formulation, large combinatorial design space, and measurable sensory, nutritional, and environmental attributes provide a tractable example of multi-objective food design. While some elements of this framework may extend to other food categories, their relevance will depend on category-specific ingredients, processing strategies, structures, characterization methods, and performance criteria.}
For burger formulation, the ingredient library shown in Figure \ref{fig03} defines this discrete design space, while ingredient quantities define its continuous counterpart. Recent work introduced a two-stage diffusion framework that integrates {\emph{multinomial diffusion}} for binary ingredient selection and {\emph{score-based diffusion}} for continuous ingredient quantification \citep{tac26b}. By learning directly from the food representations introduced in Section~3, this architecture explores formulation space while respecting the hybrid nature of food recipes. 
For burger formulation, the design space in Figure \ref{fig03} consists of 146 ingredients, corresponding to more than $2^{146} \approx 9 \times 10^{43}$ or about ninety trillion trillion trillion candidate formulations, an astronomically large design space. When trained on more than 2,200 burger recipes and used to generate one million previously unseen formulations, the model reproduced ingredient frequencies, recipe lengths, ingredient correlations, and quantity distributions with high fidelity, demonstrating that it had learned the underlying organization of burger recipes \citep{tac26}. Remarkably, random sampling rediscovered the canonical Big Mac even though the training data never contained the recipe. This result showed that the model learned {\emph{design principles}} rather than memorized individual examples. More importantly, the same framework created entirely new burgers optimized for palatability, sustainability, or nutrition. Blind sensory evaluation showed that several AI-designed burgers matched or exceeded the consumer acceptance of the Big Mac benchmark while reducing environmental impact: a mushroom-based burger achieved more than an order of magnitude lower environmental impact than the Big Mac while maintaining comparable consumer acceptance.

These advances illustrate a broader evolution in computational food design. {\emph{Large language models}} organize and communicate culinary knowledge, while {\emph{diffusion models}} transform that knowledge into optimized food formulations. Ingredients become {\emph{design variables}}, while taste, nutrition, sustainability, texture, cost, and manufacturability become {\emph{design objectives}}. Generation therefore becomes {\emph{optimization over formulation space}} rather than reproduction of existing recipes.
In practice, food formulation rarely involves a single objective. Improving nutrition may reduce consumer acceptance; minimizing greenhouse gas emissions may alter flavor or increase cost; enhancing texture may require additional processing or ingredients. Modern generative models therefore solve {\emph{multi-objective optimization}} problems that balance competing design objectives rather than maximizing a single performance metric. Instead of producing one optimal recipe, these models identify {\emph{Pareto frontiers}} 
%%%%%%%%%%%%%%%%%%%%%%%%%%%%%%%%%%%%%%%%%%%%%%%%%%%%%%%%%%%%%%%%%%%%%%%%%%%%%
\begin{marginnote}[]
\entry{pareto frontier}
{Set of optimal solutions where improving one objective degrades another. In food science, it balances taste, nutrition, sustainability, and cost.}
\end{marginnote}
%%%%%%%%%%%%%%%%%%%%%%%%%%%%%%%%%%%%%%%%%%%%%%%%%%%%%%%%%%%%%%%%%%%%%%%%%%%%%
that quantify the trade-offs among taste, nutrition, sustainability, cost, and manufacturability. Scientists, manufacturers, and consumers can then select formulations that best satisfy their priorities while making these trade-offs explicit.

{\emph{Agentic artificial intelligence}} completes this progression by transforming computational food design from recipe generation into autonomous scientific discovery. While today's generative models create candidate formulations, future agentic systems will autonomously formulate hypotheses, retrieve scientific knowledge, generate recipes, design experiments, analyze results, and iteratively refine formulations within closed-loop discovery platforms \citep{yao23}. Coupled with {\emph{world models}} that simulate food processing, fermentation, storage, and consumer response, these systems will evaluate candidate formulations before physical testing begins and continuously improve them through experimentation \citep{swanson25}. They will integrate nutritional guidelines, environmental metrics, allergen restrictions, ingredient availability, processing constraints, manufacturing limitations, and consumer preferences directly into the design process while autonomously interacting with external databases, simulation tools, and laboratory instrumentation \citep{schick23}. Rather than simply generating recipes, agentic AI will increasingly integrate the complementary capabilities introduced in Figure~\ref{fig01}—prediction, discovery, creation, knowledge integration, simulation, and autonomous action—into a unified framework for computational food design \citep{boiko23}.\\[8.pt]
Taken together, predictive and generative models establish a new computational paradigm for food formulation: {\emph{forward models}} learn how ingredients determine nutritional quality, texture, sensory perception, and shelf life, while {\emph{inverse models}} use this knowledge to discover entirely new formulations that satisfy multiple competing objectives. Predictive models estimate food properties, generative models create new formulations, foundation models organize culinary knowledge, and future world models and agentic systems will simulate and autonomously refine food designs. Together, these capabilities increasingly converge into unified AI systems for computational food design. The remaining challenge no longer lies primarily in algorithm development, but in integrating richer representations of food chemistry, mechanics, perception, manufacturing, and consumer behavior into unified design frameworks. Addressing this challenge will transform AI from a tool that predicts food properties into a scientific partner that accelerates the discovery of healthier, more sustainable, and more enjoyable foods.
%%%%%%%%%%%%%%%%%%%%%%%%%%%%%%%%%%%%%%%%%%%%%%%%%%%%%%%%%%%%%%%%%%%%%%%%%%%%
\section{AI FOR SUSTAINABLE FOODS}
%%%%%%%%%%%%%%%%%%%%%%%%%%%%%%%%%%%%%%%%%%%%%%%%%%%%%%%%%%%%%%%%%%%%%%%%%%%%
%%%%%%%%%%%%%%%%%%%%%%%%%%%%%%%%%%%%%%%%%%%%%%%%%%%%%%%%%%%%%%%%%%%%%%%%%%%%%
\begin{marginnote}[]
\entry{feature vector}
{Numerical representation that describes an object through measurable attributes. In food science, it encodes ingredients, nutrition, sustainability, or texture.}
\end{marginnote}
%%%%%%%%%%%%%%%%%%%%%%%%%%%%%%%%%%%%%%%%%%%%%%%%%%%%%%%%%%%%%%%%%%%%%%%%%%%%%
Sustainability illustrates how digital food representations transform environmental performance from a quantity that is measured after formulation into a design objective that artificial intelligence can predict, generate, and optimize.
%%%%%%%%%%%%%%%%%%%%%%%%%%%%%%%%%%%%%%%%%%%%%%%%%%%%%%%%%%%%%%%%%%%%%%%%%%%%
\subsection{Sustainability Data}
\label{sec_sustain_data}
%%%%%%%%%%%%%%%%%%%%%%%%%%%%%%%%%%%%%%%%%%%%%%%%%%%%%%%%%%%%%%%%%%%%%%%%%%%%
Artificial intelligence can optimize environmental sustainability only when the environmental impacts of food ingredients become available in structured digital representations. Sustainability data therefore represent each ingredient through a {\emph{feature vector}} 
that quantifies its environmental footprint. These descriptors typically originate from {\emph{life cycle assessment}}, which estimates the environmental impacts of agricultural products across their production, processing, transportation, and distribution chains using standardized methodologies and global producer surveys \citep{hellweg14}.
Recent years have produced several large harmonized sustainability databases that combine hundreds of life cycle assessment studies into consistent ingredient-level environmental impact estimates \citep{poore18}.
These databases enable direct comparison across thousands of foods and provide the data foundation for sustainability-aware food design \citep{clark22}. More specialized datasets have expanded this coverage to previously underrepresented food groups, including mushrooms \citep{goglio24} and aquatic foods \citep{gephart21}. Together, these resources transform environmental sustainability into a quantitative representation that AI models can learn, compare, and ultimately optimize.

In practice, most current food AI studies represent each ingredient using four complementary environmental descriptors: {\emph{land use}}, {\emph{eutrophication potential}}, {\emph{water use}}, and {\emph{greenhouse gas emissions}}. Together, these quantities capture the dominant environmental pressures associated with food production and provide a compact yet informative representation of sustainability. Many studies additionally aggregate these descriptors into a single {\emph{Environmental Impact Score}}, which simplifies comparison among ingredients and recipes while preserving their relative environmental ranking \citep{clark22}. This scalar representation provides a convenient optimization objective for predictive and generative AI models \citep{tac26}.
%%%%%%%%%%%%%%%%%%%%%%%%%%%%%%%%%%%%%%%%%%%%%%%%%%%%%%%%%%%%%%%%%%%%%%%%%%%%%
\begin{marginnote}[]
\entry{land use}
{Land area required to produce food, 
typically reported as m$^2$/year.}
\end{marginnote}
%%%%%%%%%%%%%%%%%%%%%%%%%%%%%%%%%%%%%%%%%%%%%%%%%%%%%%%%%%%%%%%%%%%%%%%%%%%%
\vspace*{-0.4cm}
%%%%%%%%%%%%%%%%%%%%%%%%%%%%%%%%%%%%%%%%%%%%%%%%%%%%%%%%%%%%%%%%%%%%%%%%%%%%%
\begin{marginnote}[]
\entry{eutrophication potential}
{Nutrient pollution potential, typically reported as kg PO$_4^{3-{\rm{eq}}}$.}
\end{marginnote}
%%%%%%%%%%%%%%%%%%%%%%%%%%%%%%%%%%%%%%%%%%%%%%%%%%%%%%%%%%%%%%%%%%%%%%%%%%%%
\vspace*{-0.4cm}
%%%%%%%%%%%%%%%%%%%%%%%%%%%%%%%%%%%%%%%%%%%%%%%%%%%%%%%%%%%%%%%%%%%%%%%%%%%%%
\begin{marginnote}[]
\entry{water use}
{Scarcity-weighted freshwater consumption, 
typically reported as L$^{\rm{eq}}$ or kL$^{\rm{eq}}$.}
\end{marginnote}
%%%%%%%%%%%%%%%%%%%%%%%%%%%%%%%%%%%%%%%%%%%%%%%%%%%%%%%%%%%%%%%%%%%%%%%%%%%%
\vspace*{-0.4cm}
%%%%%%%%%%%%%%%%%%%%%%%%%%%%%%%%%%%%%%%%%%%%%%%%%%%%%%%%%%%%%%%%%%%%%%%%%%%%%
\begin{marginnote}[]
\entry{greenhouse gas emissions}
{Climate impact expressed as carbon dioxide equivalents, 
typically reported as kg CO$_2^{\rm{eq}}$.}
\end{marginnote}
%%%%%%%%%%%%%%%%%%%%%%%%%%%%%%%%%%%%%%%%%%%%%%%%%%%%%%%%%%%%%%%%%%%%%%%%%%%%

\textcolor{black}{Like all life cycle assessment data, environmental impact estimates are not fixed material properties but context-dependent quantities \citep{chung22}. Environmental sustainability has become a central objective in modern food system design \citep{mcclements21}. The environmental footprint of an ingredient varies across regions, production systems, agricultural practices, and over time as weather conditions, water availability, and energy sources change. Processing choices further shape environmental impact through energy and water use, material efficiency, and sidestream utilization \citep{boom26}. Environmental impact estimates also depend on methodological choices, including system boundaries, functional units, and allocation methods for coproducts and multifunctional systems \citep{chung22}. Similarly, the four descriptors represent only part of the environmental landscape; biodiversity loss, soil health, pesticide use, energy demand, carbon sequestration, and ecosystem quality provide additional dimensions that will become increasingly important as sustainability databases continue to expand.}
%%%%%%%%%%%%%%%%%%%%%%%%%%%%%%%%%%%%%%%%%%%%%%%%%%%%%%%%%%%%%%%%%%%%%%%%%%%%%
\begin{marginnote}[]
\entry{life cycle assessment}
{Standardized method to quantify environmental impacts across a product's life cycle. In food science, it evaluates agricultural and food production systems.}
\end{marginnote}
%%%%%%%%%%%%%%%%%%%%%%%%%%%%%%%%%%%%%%%%%%%%%%%%%%%%%%%%%%%%%%%%%%%%%%%%%%%%
From an artificial intelligence perspective, extending these representations is straightforward: each additional environmental descriptor simply becomes another component of the ingredient feature vector. As sustainability databases become richer, AI models will naturally incorporate increasingly detailed, localized, and dynamic representations of environmental performance without changing their underlying learning framework.
%%%%%%%%%%%%%%%%%%%%%%%%%%%%%%%%%%%%%%%%%%%%%%%%%%%%%%%%%%%%%%%%%%%%%%%%%%%%%%
\begin{figure*}[h]
\includegraphics[width=1.0\textwidth]{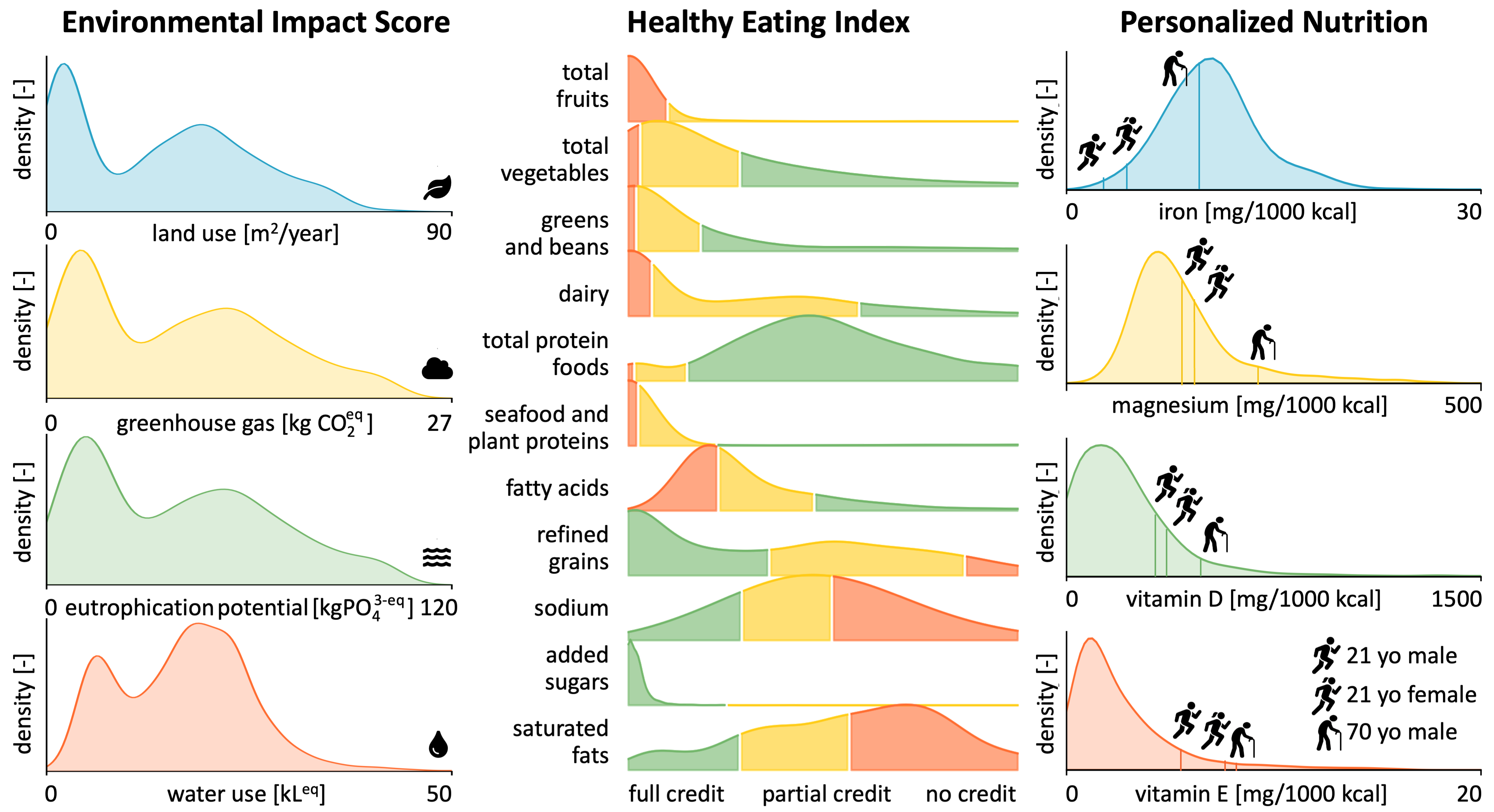}
\caption{{\bf{Generating sustainable, nutritious, and personalized foods.}} 
{\emph{Environmental Impact Factor}} for 2,216 burger recipes highlights opportunities to design more sustainable foods by reducing land use, greenhouse gas emissions, eutrophication potential, and water use (left). The same recipes span a broad range of {\emph{Healthy Eating Index}} scores and show how AI can improve nutritional quality by balancing desirable and undesirable dietary components (middle). Beyond population-level recommendations, AI can design {\emph{personalized nutrition}} by matching macro- and micronutrient compositions to individual nutritional requirements for representative demographic groups (right).}
\label{fig04}
\end{figure*} %\\[8.pt]
%%%%%%%%%%%%%%%%%%%%%%%%%%%%%%%%%%%%%%%%%%%%%%%%%%%%%%%%%%%%%%%%%%%%%%%%%%%%%%
%%%%%%%%%%%%%%%%%%%%%%%%%%%%%%%%%%%%%%%%%%%%%%%%%%%%%%%%%%%%%%%%%%%%%%%%%%%%
\subsection{Sustainability Models}
%%%%%%%%%%%%%%%%%%%%%%%%%%%%%%%%%%%%%%%%%%%%%%%%%%%%%%%%%%%%%%%%%%%%%%%%%%%%
Embedding environmental metrics into artificial intelligence models transforms sustainability from a descriptive quantity into an explicit design objective \citep{mcclements21}. Once environmental descriptors become part of the digital representation of food ingredients, AI can learn relationships between ingredient composition and environmental performance, predict sustainability before products exist, and ultimately generate entirely new formulations with reduced environmental footprints. Rather than evaluating the environmental impact of existing foods after formulation, AI can optimize ingredient combinations for reduced greenhouse gas emissions, land use, water consumption, or composite environmental scores while simultaneously satisfying constraints on nutrition, cost, manufacturability, and consumer acceptance. This progression mirrors the broader evolution of artificial intelligence from prediction toward generative and autonomous food design.
%%%%%%%%%%%%%%%%%%%%%%%%%%%%%%%%%%%%%%%%%%%%%%%%%%%%%%%%%%%%%%%%%%%%%%%%%%%%%
\begin{marginnote}[]
\entry{Environmental Impact Score}
{Composite sustainability score that combines multiple environmental indicators. In food science, it enables AI to compare and optimize food formulations.}
\end{marginnote}
%%%%%%%%%%%%%%%%%%%%%%%%%%%%%%%%%%%%%%%%%%%%%%%%%%%%%%%%%%%%%%%%%%%%%%%%%%%%

{\emph{Predictive models}} estimate greenhouse gas emissions, land occupation, freshwater use, or composite sustainability scores directly from ingredient compositions. Artificial neural networks, tree-based models, and transformer predictors learn these relationships from large environmental databases and rapidly evaluate millions of candidate formulations. These forward models establish the computational foundation for sustainability-aware food design by rapidly screening formulation space before experimental validation \citep{pennells25}.

{\emph{Foundation models}} organize and connect sustainability knowledge across recipes, ingredient databases, life cycle assessment data, nutrition, scientific literature, and consumer preferences. Large language models naturally assist ingredient substitution, summarize sustainability evidence, explain trade-offs, and support formulation decisions through natural-language interaction \citep{thomas25}. Industrial systems such as NotCo's proprietary AI platform {\emph{Giuseppe}} further illustrate how foundation models identify functional ingredient replacements that reduce reliance on animal-derived ingredients while preserving desirable product characteristics \citep{notco21}. Unlike predictive models, however, foundation models organize and communicate sustainability knowledge rather than directly optimizing quantitative environmental objectives.

{\emph{Diffusion models}} extend generative AI beyond language by directly optimizing food formulations in ingredient space. Inspired by non-equilibrium diffusion processes, these models iteratively refine noisy ingredient compositions toward recipes that satisfy multiple competing objectives simultaneously. Unlike large language models, diffusion models optimize directly in formulation space rather than text space, making it straightforward to incorporate environmental impact as an explicit design criterion during recipe generation. A recent study combined multinomial diffusion with score-based diffusion to generate sustainable burger formulations from more than 2,200 recipes \citep{tac26b}. Using the ingredient-level environmental representations introduced in Section \ref{sec_sustain_data}, the model optimized the Environmental Impact Score directly during recipe generation. Figure~\ref{fig04} illustrates the broad distribution of environmental impacts across the recipe space and highlights the opportunity for AI to discover novel sustainable formulations. The resulting beef--mushroom blend and fully mushroom-based burger reduced environmental impact by approximately one order of magnitude compared with McDonald's Big Mac while maintaining desirable nutritional quality and consumer acceptance \citep{tac26}. \\[8.pt]
Together, these approaches mirror the progression of AI capabilities introduced in Figure~\ref{fig01}. {\emph{Predictive models}} predict environmental performance from ingredient composition. {\emph{Foundation models}} know and organize relationships among ingredients, recipes, sustainability metrics, and scientific knowledge. {\emph{Diffusion models}} create new formulations by optimizing multiple environmental and consumer objectives simultaneously. {\emph{World models}} will simulate how sustainability evolves during formulation, processing, fermentation, storage, distribution, and consumption. {\emph{Agentic AI}} will autonomously design experiments, coordinate simulations and laboratory measurements, evaluate environmental outcomes, and iteratively refine formulations through closed-loop discovery workflows. Sustainability will therefore evolve from a quantity that AI predicts into a property that AI discovers, creates, understands, simulates, and ultimately realizes through autonomous food innovation.
%%%%%%%%%%%%%%%%%%%%%%%%%%%%%%%%%%%%%%%%%%%%%%%%%%%%%%%%%%%%%%%%%%%%%%%%%%%%
\section{AI FOR NUTRITIOUS FOODS}
%%%%%%%%%%%%%%%%%%%%%%%%%%%%%%%%%%%%%%%%%%%%%%%%%%%%%%%%%%%%%%%%%%%%%%%%%%%%
Nutrition illustrates how digital food representations transform dietary quality from a quantity that is evaluated after formulation into a design objective that artificial intelligence can predict, generate, and personalize.
%%%%%%%%%%%%%%%%%%%%%%%%%%%%%%%%%%%%%%%%%%%%%%%%%%%%%%%%%%%%%%%%%%%%%%%%%%%%
\subsection{Nutrition Data}
%%%%%%%%%%%%%%%%%%%%%%%%%%%%%%%%%%%%%%%%%%%%%%%%%%%%%%%%%%%%%%%%%%%%%%%%%%%%
Artificial intelligence can optimize nutritional quality only when the nutritional properties of foods become available in structured digital representations. Nutrition data therefore represent every ingredient or recipe through a {\emph{feature vector}} that quantifies its nutritional composition and dietary quality. These descriptors originate from food composition databases, nutrient profiling systems, and dietary recommendations that translate nutritional science into computationally accessible representations.

The foundation of most nutrition-aware AI models is ingredient-level nutrient composition. {\emph{USDA FoodData Central}} provides nutrient profiles for more than 300,000 foods and serves as the primary nutritional reference database in the United States \citep{ahuja20}. Complementary resources such as the {\emph{Food Patterns Equivalents Database}} translate foods into standardized food-group equivalents and enable evaluation against dietary guidelines rather than individual nutrients alone \citep{bowman20}. Together, these resources provide the data foundation for nutrition-aware food design.

\begin{figure*}[h]
\includegraphics[width=1.0\textwidth]{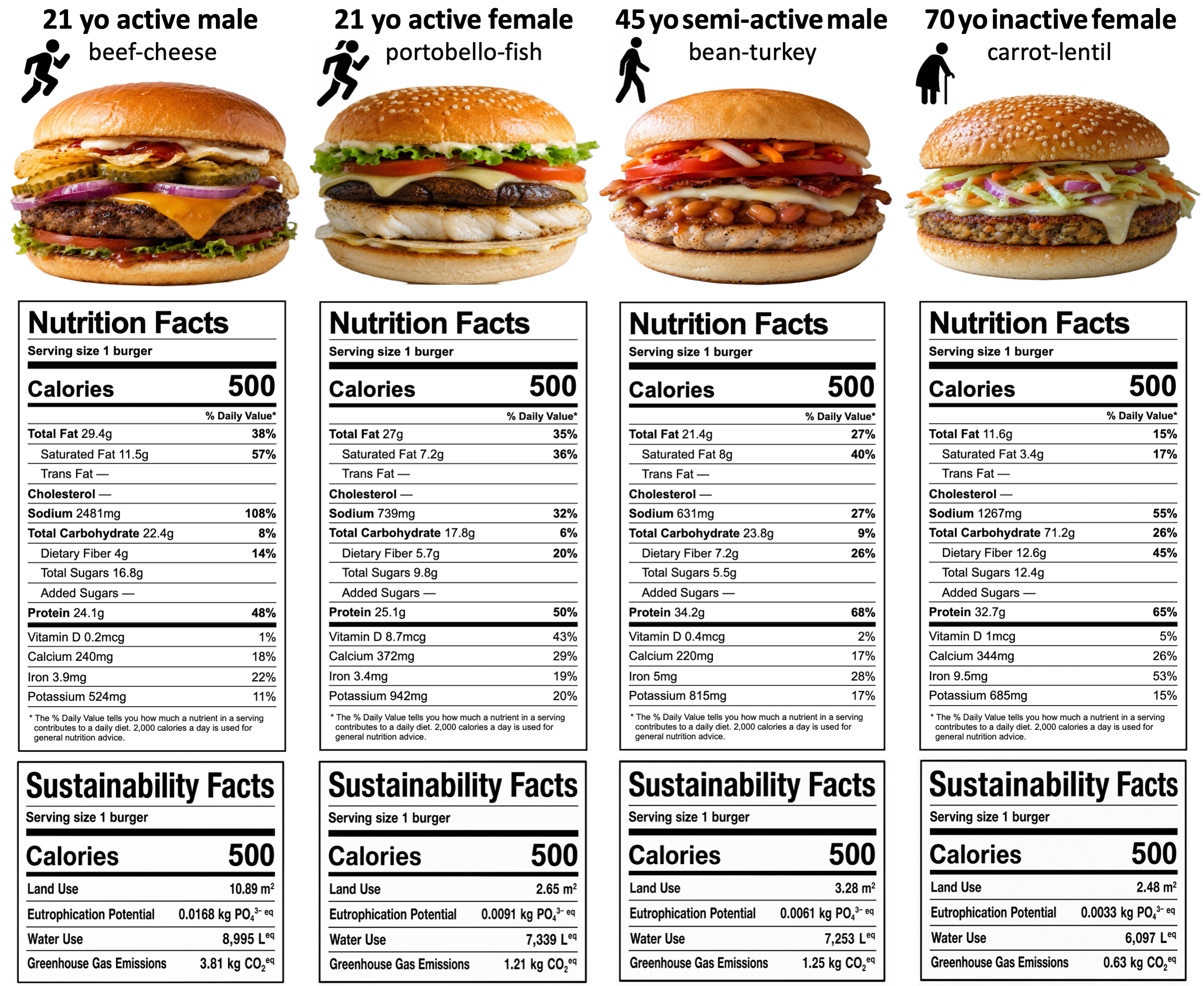}
\caption{{\bf{Personalized computational food design. }} 
Generative AI creates individualized burger formulations tailored to age, biological sex, and physical activity while simultaneously optimizing multiple competing objectives. Each recipe combines personalized nutritional profiles with quantified environmental impacts, illustrating how AI can integrate nutrition and sustainability within a unified computational design framework.
Personalized burgers were generated with the interactive {\emph{AI4Burgers}} platform \citep{tac26c}.}
\label{fig05}
\end{figure*}

In practice, nutritional quality is commonly represented through nutrient profiling systems that condense high-dimensional nutrient information into compact quantitative descriptors. The {\emph{Healthy Eating Index}} evaluates adherence to the Dietary Guidelines for Americans \citep{krebs18}. {\emph{Nutri-Score}} and the {\emph{Health Star Rating}} instead summarize nutritional quality by balancing nutrients to encourage against nutrients to limit \citep{julia18}. These scalar representations provide convenient optimization objectives for predictive and generative AI models. Beyond population-wide recommendations \citep{mainardi19}, 
personalized nutrition scores 
%%%%%%%%%%%%%%%%%%%%%%%%%%%%%%%%%%%%%%%%%%%%%%%%%%%%%%%%%%%%%%%%%%%%%%%%%%%%%
\begin{marginnote}[]
\entry{personalized nutrition}
{Nutritional approach tailored to individual biological and lifestyle characteristics. In food science, it guides AI-driven personalized food design.}
\end{marginnote}
%%%%%%%%%%%%%%%%%%%%%%%%%%%%%%%%%%%%%%%%%%%%%%%%%%%%%%%%%%%%%%%%%%%%%%%%%%%%%
incorporate age, biological sex, body size, physical activity, and other individual characteristics 
to quantify how closely foods satisfy personal nutritional requirements \citep{zeisel20}. Figure~\ref{fig05} illustrates this concept through AI-generated burger formulations tailored to different individuals.

Like sustainability data, nutritional representations continue to evolve \citep{mcclements24a}. Existing scoring systems capture only selected aspects of dietary quality and rarely account for food structure, nutrient bioavailability, glycemic response, gut microbiome interactions, or individual metabolic variability \citep{kok23}. From an artificial intelligence perspective, extending nutritional representations is straightforward: each additional nutritional descriptor simply becomes another component of the ingredient or recipe feature vector. As nutrition databases become richer and increasingly personalized, AI models will naturally incorporate more comprehensive, dynamic, and individualized representations of human nutrition without changing their underlying learning framework.
%%%%%%%%%%%%%%%%%%%%%%%%%%%%%%%%%%%%%%%%%%%%%%%%%%%%%%%%%%%%%%%%%%%%%%%%%%%%%%
 %\\[8.pt]
%%%%%%%%%%%%%%%%%%%%%%%%%%%%%%%%%%%%%%%%%%%%%%%%%%%%%%%%%%%%%%%%%%%%%%%%%%%%%%
%%%%%%%%%%%%%%%%%%%%%%%%%%%%%%%%%%%%%%%%%%%%%%%%%%%%%%%%%%%%%%%%%%%%%%%%%%%%
\subsection{Nutrition Models}
%%%%%%%%%%%%%%%%%%%%%%%%%%%%%%%%%%%%%%%%%%%%%%%%%%%%%%%%%%%%%%%%%%%%%%%%%%%%
Embedding nutritional representations into artificial intelligence models transforms nutrition from a descriptive quantity into an explicit design objective. Once nutritional descriptors become part of the digital representation of food ingredients, AI can learn relationships between ingredient composition and dietary quality, predict nutritional performance before products exist, and ultimately generate entirely new formulations tailored to specific nutritional goals. Rather than evaluating existing foods after formulation, AI can optimize recipes directly for nutritional quality while simultaneously satisfying constraints on sensory properties, sustainability, cost, manufacturability, and consumer acceptance. This progression mirrors the broader evolution of artificial intelligence from prediction toward generative and personalized food design.

{\emph{Predictive models}} can estimate nutritional quality directly from ingredient composition. Machine learning models learn relationships among ingredients, recipes, and dietary guidelines to predict nutritional properties and support nutrition-aware food reformulation \citep{papastratis24}. These forward models establish the computational foundation for nutrition-aware food design by rapidly screening formulation space before experimental validation.

{\emph{Foundation models}} organize and connect nutritional knowledge across ingredients, recipes, dietary guidelines, nutrient databases, biomedical literature, and consumer preferences. Large language models retrieve nutritional information, explain dietary recommendations, support meal planning, and assist recipe development through natural-language interaction \citep{papastratis24}. Domain-specific food foundation models further strengthen these capabilities by integrating large-scale culinary and nutrition corpora into unified representations for nutritional reasoning \citep{zhou25}. Unlike predictive models, however, foundation models organize and communicate nutritional knowledge rather than directly optimizing quantitative nutritional objectives.

{\emph{Generative models}} extend nutritional optimization beyond recipe modification by directly generating new food formulations in ingredient space. Multi-objective optimization naturally balances competing design criteria, including nutrient density, protein quality, caloric content, ingredient cost, sustainability, and consumer preference. A recent study demonstrated this concept by optimizing burger formulations for the Healthy Eating Index while simultaneously maximizing sensory quality and minimizing environmental impact through multinomial diffusion models \citep{tac26}. Unlike large language models, diffusion models optimize directly in formulation space, making it straightforward to incorporate nutritional quality as an explicit objective during recipe generation.

{\emph{Personalized nutrition models}} further extend this framework from population-level recommendations to individualized food design. Nutritional requirements depend on age, biological sex, body size, physical activity, physiological state, health status, food allergies, and dietary preferences, and growing evidence suggests that individuals respond differently to identical foods \citep{zeevi15}. Recent work demonstrated this concept by generating personalized burger recipes tailored to an individual's age, height, weight, biological sex, and physical activity level while simultaneously maintaining sensory quality and sustainability \citep{tac26}. Figure~\ref{fig05} illustrates four personalized formulations generated for individuals with different ages, biological sexes, and activity levels. As wearable sensors, digital health records, continuous metabolic monitoring, and multi-omics data continue to expand, AI will increasingly generate foods designed for the nutritional requirements of individual consumers rather than average populations, although substantial scientific and clinical validation remains necessary before personalized nutrition can realize its full potential \citep{ordovas18}.\\[8.pt]
Together, these approaches illustrate the progression introduced in Figure~\ref{fig01}. \textcolor{black}{{\emph{Predictive models}} estimate nutritional properties from ingredient composition and other food representations.} {\emph{Foundation models}} know and organize relationships among ingredients, recipes, dietary guidelines, nutrient databases, and biomedical knowledge. {\emph{Generative models}} create healthier food formulations by optimizing multiple nutritional objectives simultaneously. {\emph{Personalized nutrition models}} adapt these formulations to the nutritional requirements of individual consumers. {\emph{World models}} will simulate long-term physiological responses to foods, while {\emph{Agentic AI}} will autonomously design, evaluate, and refine personalized nutritional interventions through closed-loop experimentation. Nutrition will therefore evolve from a quantity that AI predicts into a property that AI discovers, creates, personalizes, simulates, and ultimately realizes through autonomous nutritional design.
%%%%%%%%%%%%%%%%%%%%%%%%%%%%%%%%%%%%%%%%%%%%%%%%%%%%%%%%%%%%%%%%%%%%%%%%%%%%
\section{TOWARD THE GENERATIVE SCIENCE OF FOOD FORMULATION}
%OPPORTUNITIES AND OPEN QUESTIONS
%%%%%%%%%%%%%%%%%%%%%%%%%%%%%%%%%%%%%%%%%%%%%%%%%%%%%%%%%%%%%%%%%%%%%%%%%%%%
The rapid progress of artificial intelligence has demonstrated that computational food design is becoming scientifically feasible. The preceding sections define the foundations of this emerging discipline. \textcolor{black}{The following scientific challenges will shape the generative science of food formulation.}\\[6.pt]
%%%%%%%%%%%%%%%%%%%%%%%%%%%%%%%%%%%%%%%%%%%%%%%%%%%%%%%%%%%%%%%%%%%%%%%%%%%%%
%\begin{issues}[FUTURE ISSUES]
%%%%%%%%%%%%%%%%%%%%%%%%%%%%%%%%%%%%%%%%%%%%%%%%%%%%%%%%%%%%%%%%%%%%%%%%%%%%%
%\begin{enumerate}
%\item Open benchmarks enable scientific progress.
%\item Multimodal representations connect food science.
%\item Physics grounds AI in reality.
%\item Trust requires transparency.
%\item Food AI should remain an open science.
%\end{enumerate}
%\end{issues}
%%%%%%%%%%%%%%%%%%%%%%%%%%%%%%%%%%%%%%%%%%%%%%%%%%%%%%%%%%%%%%%%%%%%%%%%%%%%
\noindent {\bf{Open benchmarks enable scientific progress.}} Predictive and generative models can only be compared when they learn from shared datasets and standardized evaluation protocols. Unlike computer vision and natural language processing, food science still lacks widely adopted benchmark datasets that combine ingredients, nutritional composition, sensory perception, texture, sustainability, processing, and shelf-life measurements within unified experimental frameworks. Establishing open, multimodal benchmarks will improve reproducibility, enable fair comparison among competing algorithms, and accelerate methodological progress across academia and industry.\\[6.pt]
{\bf{Multimodal representations connect food science.}} Food formulation requires integrating heterogeneous sources of information that span chemistry, nutrition, mechanics, sensory science, manufacturing, and consumer behavior. Current databases remain fragmented across disciplines and frequently use incompatible standards and ontologies. The next generation of food foundation models should learn unified representations that connect these complementary data modalities. Such models could reason simultaneously across recipes, molecular composition, rheological behavior, nutritional quality, environmental impact, and scientific literature, creating a shared computational language for food science.\\[6.pt]
\textcolor{black}{{\bf{Physics grounds AI in reality.}} 
Statistical learning alone cannot guarantee physically meaningful formulations. Modern food AI increasingly combines statistical learning with mechanistic food physics. Embedding food physics, rheology, transport phenomena, thermodynamics, and oral processing into predictive and generative models improves robustness, extrapolation, and scientific interpretability. Rather than replacing mechanistic understanding, hybrid computational frameworks combine first-principles models with data-driven learning to generate foods that are not only desirable but also physically plausible and manufacturable.} \\[6.pt]
{\bf{Trust requires transparency.}} As AI increasingly influences nutritional recommendations, sustainability assessments, and food design, transparency becomes essential. Scientists, regulators, manufacturers, and consumers must understand how algorithms arrive at their recommendations and which assumptions underlie their predictions. Explainable models, uncertainty quantification, standardized reporting practices, and transparent validation protocols will therefore become critical components of trustworthy AI for food science. \textcolor{black}{When proprietary algorithms prevent full transparency, independent validation, auditable outputs, uncertainty estimates, and explicit disclosure of model assumptions become even more important.}\\[6.pt]
{\bf{Food AI should remain an open science.}} Food directly affects human health, environmental sustainability, and global food security. The scientific infrastructure that supports AI-driven food design should therefore remain as open and accessible as possible. \textcolor{black}{If data, models, and computational infrastructure remain proprietary, there is a risk that AI could concentrate food innovation within a small number of large companies and widen existing disparities in access to technology. Open datasets, open-source models, interoperable software, and shared evaluation standards can counter this concentration by lowering barriers to participation and broadening access to innovation. Such openness can help ensure that advances benefit researchers, industry, policymakers, and consumers rather than remaining confined to a small number of proprietary platforms.}\\[6.pt]
Together, these challenges extend beyond artificial intelligence itself. They define the scientific infrastructure required for computational food design to evolve from isolated demonstrations into a mature design discipline. Addressing these challenges will establish the scientific infrastructure that allows artificial intelligence not only to predict food properties, but also to discover scientific principles, create new formulations, organize knowledge, simulate food systems, and ultimately act as an autonomous partner in food innovation.
%%%%%%%%%%%%%%%%%%%%%%%%%%%%%%%%%%%%%%%%%%%%%%%%%%%%%%%%%%%%%%%%%%%%%%%%%%%%%
%\begin{summary}[SUMMARY POINTS]
%%%%%%%%%%%%%%%%%%%%%%%%%%%%%%%%%%%%%%%%%%%%%%%%%%%%%%%%%%%%%%%%%%%%%%%%%%%%%
%\begin{enumerate}
%\item Predictive AI estimates food properties; generative AI creates new formulations.
%\item Food formulations become design objects that AI generates and optimizes.
%\item Modern AI balances taste, nutrition, sustainability, cost, and manufacturability.
%\item World models and agentic AI automate discovery, experiments and optimization.
%\item Food formulation is becoming a generative science.
%\end{enumerate}
%\end{summary}
%%%%%%%%%%%%%%%%%%%%%%%%%%%%%%%%%%%%%%%%%%%%%%%%%%%%%%%%%%%%%%%%%%%%%%%%%%%%%
%%%%%%%%%%%%%%%%%%%%%%%%%%%%%%%%%%%%%%%%%%%%%%%%%%%%%%%%%%%%%%%%%%%%%%%%%%%%
\section{CONCLUSION}
%%%%%%%%%%%%%%%%%%%%%%%%%%%%%%%%%%%%%%%%%%%%%%%%%%%%%%%%%%%%%%%%%%%%%%%%%%%%
\textcolor{black}{The long-term goal is not separate AI systems for flavor, nutrition, sustainability, processing, or safety, but a unified computational framework that simultaneously optimizes multiple design objectives while satisfying essential design constraints.} Artificial intelligence is changing far more than the tools we use to formulate foods. It is changing the way we think about food formulation itself. By combining digital food representations with predictive, generative, and increasingly autonomous AI systems, food formulation is shifting from empirical recipe development toward computational design.
In this review, we present a unified framework for this transformation. We describe the evolution of artificial intelligence from prediction to discovery, creation, knowledge integration, simulation, and autonomous action, and show how these capabilities operate on structured representations of ingredients, nutrition, flavor, texture, sustainability, and shelf life. {\emph{Once these food properties become digital representations, they become quantities that artificial intelligence can predict, understand, generate, and optimize.}} Sustainability and nutrition illustrate this broader paradigm, but the same principles naturally extend to many other objectives in food design.
The next advances will depend not only on better algorithms, but also on open benchmarks, multimodal food representations, mechanistic models that embed food physics, transparent and trustworthy AI systems, and an open scientific infrastructure that enables reproducible and collaborative research. Together, these foundations will establish computational food design as a rigorous scientific discipline. The next era of food innovation will not simply discover better foods--it will design them. The convergence of digital food representations, mechanistic understanding, and modern artificial intelligence is transforming food science from an empirical discipline into a predictive, generative, and increasingly autonomous design science. The generative science of food formulation has begun.
%%%%%%%%%%%%%%%%%%%%%%%%%%%%%%%%%%%%%%%%%%%%%%%%%%%%%%%%%%%%%%%%%%%%%%%%%%%%
\section*{DISCLOSURE STATEMENT}
%%%%%%%%%%%%%%%%%%%%%%%%%%%%%%%%%%%%%%%%%%%%%%%%%%%%%%%%%%%%%%%%%%%%%%%%%%%%
The authors are not aware of any affiliations, memberships, funding, or financial holdings that might be perceived as affecting the objectivity of this review. 
%%%%%%%%%%%%%%%%%%%%%%%%%%%%%%%%%%%%%%%%%%%%%%%%%%%%%%%%%%%%%%%%%%%%%%%%%%%%
\section*{ACKNOWLEDGMENTS}
%%%%%%%%%%%%%%%%%%%%%%%%%%%%%%%%%%%%%%%%%%%%%%%%%%%%%%%%%%%%%%%%%%%%%%%%%%%%
\noindent
This research was supported by
the Schmidt Science Fellowship 
in partnership with the Rhodes Trust to Vahidullah Tac, and by
the Stanford Bio-X Snack Grant 2025,
the Stanford SDSS Accelerator Grant 2025,
the NSF CMMI Award 2320933, and
the ERC Advanced Grant 101141626 to Ellen Kuhl.
%%%%%%%%%%%%%%%%%%%%%%%%%%%%%%%%%%%%%%%%%%%%%%%%%%%%%%%%%%%%%%%%%%%%%%%%%%%%

%%%%%%%%%%%%%%%%%%%%%%%%%%%%%%%%%%%%%%%%%%%%%%%%%%%%%%%%%%%%%%%%%%%%%%%%%%%%
%%%%%%%%%%%%%%%%%%%%%%%%%%%%%%%%%%%%%%%%%%%%%%%%%%%%%%%%%%%%%%%%%%%%%%%%%%%%
\end{document}